\documentclass[preprint,showpacs,preprintnumbers,amsmath,amssymb,nofootinbib]{revtex4}

\usepackage{graphicx}
\usepackage{dcolumn}
\usepackage{bm}

\catcode`\@=11

\def\lsim{\mathrel{\mathpalette\@versim<}}

\def\gsim{\mathrel{\mathpalette\@versim>}}
\def\@versim#1#2{\vcenter{\offinterlineskip
\ialign{ \m@th#1\hfil##\hfil \crcr#2\crcr\sim\crcr } }}
\catcode`\@=12

\newcommand{\be}{\begin{eqnarray}}
\newcommand{\ee}{\end{eqnarray}}
\newcommand{\nn}{\nonumber}

\begin{document}

\pagestyle{empty}

 \preprint{KANAZAWA-04-16}

\title{Dihedral Families of Quarks, Leptons and Higgs Bosons}

\vspace{2cm}

\author{K.~S.~Babu$^{1}$}

\author{Jisuke Kubo$^{2}$}

\vspace{1cm}

\affiliation{
$^{1}$Department of Physics,
Oklahoma State University,
Stillwater, OK 74078, USA\\
$^{2}$Institute for Theoretical Physics, 
Kanazawa University, Kanazawa 920-1192, Japan
 }

\vspace{3cm}
\begin{abstract}
We consider finite groups of small order for family symmetry.
It is found that the binary dihedral  group $Q_6$,
along with the assumption that the Higgs sector is of type II,
predicts mass matrix of a nearest neighbor interaction type   for quarks and leptons.
We present a supersymmetric  model based on $Q_6$ with
spontaneously   induced CP phases.
The quark  sector contains 8 real parameters with one independent phase
to describe  the quark masses  and  their mixing. 
Predictions  in the $|V_{ub}|-\bar{\eta}$, $|V_{ub}|-\sin 2 \beta(\phi_1)$
and $|V_{ub}|-|V_{td}/V_{ts}|$ planes are given.
The  lepton sector contains also 9 parameters.
A normal as well as an inverted spectrum of neutrino masses is possible,  and we 
compute $V_{e3}$.
We find that 
$|V_{e3}|^2 >  10^{-4} $ in the case of a normal spectrum, and
$|V_{e3}|^2  >8 \times 10^{-4}$ in the case of an inverted spectrum.
It is also found that  $Q_6$ symmetry forbids all Baryon number violating terms of $d=4$,
and the contributions to EDMs from the A terms vanish in this model.

\end{abstract}


\maketitle

\newpage
\renewcommand{\baselinestretch}{1.0}
\pagestyle{plain}
\pagenumbering{arabic}
\setcounter{footnote}{0}

\section{Introduction}
The gauge interactions of the standard model (SM)
respect $U_L(3) \times U_R(3)$ family symmetry
in both  the leptonic and quark sectors.
 It is the Yukawa sector of the SM that breaks this family symmetry,
and  is responsible for the generation of  the 
 lepton
and quark masses and their mixing.
If no condition is imposed, there are
$3\times 3\times 2 \times 2
-8=28$ real free parameters in the quark sector alone,
where $8$ is the number of phases that can be
absorbed into the phases of the quark fields.
Of $28$ only $10$ parameters are physical parameters in the SM.
That is, there are
$18$  redundant parameters in that sector.
The presence of  redundant parameters 
is not related to a symmetry in the SM. 
Even if they are set equal to zero at some energy scale, 
they will appear at different scales.
These redundant parameters may become physical
parameters 
when going beyond the SM, and, moreover, they can induce
flavor changing neutral currents (FCNCs) and CP
violating phenomena that are absent or strongly suppressed
in the SM.
Since the SM can not control the redundant parameters,
the size of the new FCNCs and CP violating phases
 may be unacceptably  large unless
 there is some symmetry, or 
 one fine tunes  their values.
 This is a flavor problem that can occur
 when going beyond the SM,
 and the most familiar case is the minimal supersymmetric
 standard model (MSSM) \cite{susy}.

 Reducing
 phenomenologically  the number of the free parameters
 in the mass matrices of quarks and leptons
 had started
 already decades ago. One of the successful Ans\" atze 
 for the quark mass matrices,
 first proposed by Weinberg \cite{weinberg,wilczek}and then extended by  
 Fritzsch \cite{fritzsch1},  is
of a nearest neighbor interaction (NNI) type \footnote{See 
\cite{fritzsch2} for review.}
\cite{branco1}-\cite{fukugita1}:
 \be
 M &=&
  \left( \begin{array}{ccc}
0  & C &0 \\
\pm C & 0 &B\\
0 &  B'  &A\\
\end{array}\right).
\label{nni}
 \ee
The complex parameters
$B, B', C$ and $A$  for each of the up and down quark sectors can
be made real by an appropriate phase rotation on the quark fields,
and as a consequence,  there are only $8$ real free parameters with two
independent phases. 
The Ansatz (\ref{nni}) can successfully reproduce
the quark masses and the Cabibbo-Kobayashi-Maskawa (CKM) mixing matrix
$V_{\rm CKM}$ \cite{dutta1,fritzsch2}.
It has been also realized that the  Ansatz (\ref{nni}) can be used  in the 
leptonic sector, too \cite{davies1,babu1, achiman1, xing1, fukugita1}.
Therefore, the Ansatz (\ref{nni})  is appropriate for  unification, especially 
for the Pati-Salam type unification \cite{patisalam}, 
in which the left-handed and right-handed fermion
families can be separately unified.

It is known \cite{branco1} that  within the SM, 
any mass matrix for both up and down quarks
can be simultaneously brought,  without changing physics, to the from 
(\ref{nni}) with $|M_{12}| \neq |M_{21}| $.
However, beyond the SM, this is no longer true, and
to obtain (\ref{nni})  beyond the SM
even with $|M_{12}| \neq |M_{21}|$, some principle 
should be  required. 
In this paper we are motivated by a desire to derive the form (\ref{nni}) 
solely from a symmetry principle.
To be definite, we assume that the responsible symmetry
is (A) based on a nonabelian discrete 
group, and (B)
 only spontaneously broken.
 As we will find, (i) the smallest group
that satisfies our assumptions is $Q_6$, a 
binary dihedral  group
with $12$ elements, and (ii)  the Higgs sector of the SM
has to be so extended that the up- and down-type right-handed fermion families
 couple to their own $SU(2)_L$ Higgs  doublets (type II Higgs).
So, the Higgs sector
of the MSSM  fits  the desired Higgs structure.
Therefore, we are naturally led to  consider
a supersymmetric extension of the SM based
 on $Q_6$, as we will do in this paper.
[ Frampton and Kephart \cite{frampton1}
 came to $Q_6$, but from different 
 reasons \footnote{
One of the first papers
on discrete symmetries  are
\cite{pakvasa1,harari,derman,wyler,segre,sato,yamanaka,koide2}.
Phenomenologically
viable models based on 
nonabelian discrete flavor symmetries $A_{4}, S_{3}, D_{4}$ 
and $Q_4$,
which can partly explain  the flavor 
structure of quarks and leptons such as large neutrino mixing,
 have been recently constructed, respectively, in
 \cite{ma1,babu2,ma4},  \cite{ma0,kubo1,kubo2,ma2},  \cite{grimus2,grimus3},  
and \cite{frigerio}. In \cite{hall1,hamaguchi1,babu3,kobayashi1,choi1},
nonabelian discrete symmetries have been used to soften the SUSY flavor problem.}.]
 We will also discuss other important consequences of 
 the supersymmetric $Q_6$ model such as 
 a solution to the SUSY flavor problem  \cite{gabbiani1}.
We find  that 
 CP can be spontaneously broken, and thanks to $Q_6$
 a  phase alignment for each A term  (trilinear coupling of 
bosonic superpartners)
 with the corresponding Yukawa term occurs. 
Note that the misalignment of the phases appearing in the Yukawa 
and  A terms is 
the origin of a large contribution to the electric dipole moments (EDM)
of neutron etc \cite{fcnc-edm}. 
We also find that $Q_6$ can forbid all the Baryon number violating
$d=3$ and $4$ operators, and allows only one
$R$-parity violating operator with $d  \leq 4$.

After we discuss group theory on the dihedral groups $D_N$
and the binary dihedral groups $Q_N$ \cite{frampton1,frampton2} in Sect. II,
we consider a supersymmetric extension of the SM based on $Q_6$
in Sect. III.
There we discuss the quark sector, lepton sector and Higgs sector,
separately. We make  predictions  in the $|V_{ub}|-\bar{\eta},
|V_{ub}|-\sin2\beta(\phi_1)$ and $ |V_{ub}|-V_{td}/V_{ts}|$ planes
as well as on the average neutrino mass $<m_{ee}>$
in neutrinoless double $\beta$ decay and the Dirac phase $\delta_{CP}$
in the neutrino mixing.
In the last section, we briefly discuss the
anomalies of discrete symmetries \cite{banks,ibanez,babu4} introduced for the model,
$R$ parity violating operators and the 
SUSY flavor problem, respectively.

\section{Finite groups $D_N$ and $Q_N$}
 \subsection{Definitions}
 The group presentation for
the dihedral groups $D_N$  is given by
\be
\{A_{D_N},B_D; (A_{D_N})^N = B_D^2 = E,
~B_D^{-1}A_{D_N}B_D = A_{D_N}^{-1}\},
\ee
and
\be
\{A_{Q_N},B_Q; (A_{Q_N})^{N} =E,  B_Q^2 = (A_{Q_N})^{N/2},
B_Q^{-1}A_{Q_N} B_Q= A_{Q_N} ^{-1}\}
\ee
for the binary dihedral  group $Q_N$,
where $E$ is the identity element.
$2N$ is the order of 
group (the number of the group elements).
For the binary dihedral group $Q_N$, $N$ should be even starting with $4$,
while $N$ for $D_N$ starts with $3$.
The $2N$ group elements are:
\be
{\cal G} =\{E, A, (A)^2, \dots, ( A)^{N-1}, B, AB, (A)^2 B, \dots, (A)^{N-1} B\}
\label{elements}
\ee
both for $D_N$ and $Q_N$.
Using the property that
the product $(A^m B)(A^n B)$ with $ m,n=0,\dots,N-1$ can always be brought to
one of the elements of ${\cal G}$, 
one can easily see that they form a group.
A two-dimensional representation of $A$ and $B$
is given by
\begin{eqnarray}
A_{D_N} &=&A_{Q_N} = 
\left(\begin{array}{cc}
\cos \phi_N & \sin\phi_N  \\
-\sin \phi_N & \cos\phi_N  \end{array}\right)~\mbox{with}~\phi_N=2\pi/N,
\label{AD}\\
 B_D &=& \left(\begin{array}{cc}1 & 0 \\ 0 & -1 \end{array}\right)~~\mbox{for}~D_N,~
  B_Q = \left(\begin{array}{cc} i & 0 \\ 0 & -i \end{array}\right)~~\mbox{for}~Q_N.
  \label{BD}
\end{eqnarray}
Note that 
$\det A_{Q_N}=\det B_{Q_N}=1$, 
implying that $Q_N$ is a subgroup of $SU(2)$.
It follows that the dihedral group is a subgroup of $SO(3)$, which one sees
if one  embeds $A_{D_N}$ and $B_D$ into $3\times 3$ matrices
\be
A_{D_N} &\to & 
\left(\begin{array}{ccc}\cos \phi_N & \sin\phi_N &0 \\ -\sin \phi_N & \cos
\phi_N &0 \\ 0 & 0 &1 \end{array}\right),
B_D \to 
 \left(\begin{array}{ccc}1 & 0 &0 \\ 0 & -1& 0 \\ 0 & 0 &-1  \end{array}\right).
 \ee
It also follows that $D_N$ has only real representations,
while $Q_N$ can have real as well as pseudo-real representations \cite{frampton1,frampton2}.
However, the smallest binary dihedral group that contains
both  real and pseudo-real nonsinglet representations is $Q_6$, because
$Q_4$ has only pseudo-real  nonsinglet representations.
Note that the irreducible representations (irreps) of $D_N$ and $Q_N$ are
either one- or two-dimensional.

$Q_N$ is the ``double-covering group'' of $D_N$ in the following sense.
Consider the matrices of $D_{N/2}$, i.e.,
$A_{D_{N/2}}$ and $B_D$, and define
\be
\tilde{A}_{Q_N} &=& A_{D_{N/2}}, ~\tilde{B}_Q=B_D.
\ee
Note that $\tilde{A}_{Q_N}$ have exactly the same properties as 
$A_{Q_N}$. Therefore, the set
\be
\{E, \tilde{A}_{Q_N}, (\tilde{A}_{Q_N})^2, 
\dots, ( \tilde{A}_{Q_N})^{N-1}, \tilde{B}_Q,
\tilde{A}_{Q_N}\tilde{B}_Q, 
(\tilde{A}_{Q_N})^2 \tilde{B}_Q, \dots, (\tilde{A}_{Q_N})^{N-1} \tilde{B}_Q\}
\label{elements2}
\ee
is a set of $Q_N$ elements.
Since however $(\tilde{A}_{Q_N})^{N/2}=
( A_{D_{N/2}})^{N/2}=E$ by definition, 
 the $D_{N/2}$ elements appear twice in 
(\ref{elements2}).

\subsection{$Q_6$ group theory}
Before we consider  a concrete model, we briefly
outline the basic reasons why $Q_6$.
Let us find out what   
symmetry can explain the right-upper
$2 \times 2$ block of the mass matrix (\ref{nni}). 
Clearly, no abelian symmetry can explain it,
whatever the Higgs structure is.
In fact, that block with
$M_{12}=-M_{21}=C$, is invariant under
$SU(2)$. It is therefore invariant under 
its subgroups, too, in particular under  $Q_{N}$.
Similarly, one can interpret that the $2 \times 2$ block
is invariant under $O(2)$. Note that $O(2)$ is not
abelian ($2 \times 2$ orthogonal matrices with
$\det =\pm 1$ do not always commute with each other),
and that $O(2)$ allows two different one-dimensional
representations ${\bf 1}$ and ${\bf 1}'$, where only ${\bf 1}$ 
is the true singlet. The diagonal entries of the $2 \times 2$ block
can be forbidden if  ${\bf 1}'$ is assigned to the responsible Higgs.
In fact, $O(2)$ may be regarded as  a subgroup
of $SO(3)$, and in this case the nonabelian finite subgroups  are
the dihedral groups, $D_N$.
It turns out that to explain the whole structure of the mass matrix
(\ref{nni}), we need to have real as well as pseudo-real representations.

 The smallest group  that contains both types of representations is $Q_6$, which is 
 the double-covering group of $S_3 \sim D_3$.
 The irreps of $Q_6$ are ${\bf 2}, ~ {\bf
2'},~{\bf 1},~{\bf 1'},~{\bf 1''},~{\bf 1'''}$, where the ${\bf
2}$ is pseudo-real, while ${\bf 2'}$ is real.  ${\bf 1},~ {\bf 1'}$
are real representations, while ${\bf 1''},~{\bf 1'''}$ are
complex conjugate to each other.   The group multiplication 
rules are given as follows \cite{frampton1,frampton2}:
\begin{eqnarray}
{\bf 1'} \times {\bf 1'} &=& {\bf 1},~~{\bf 1''} \times 
{\bf 1''} = {\bf 1'},~~{\bf 1'''} \times {\bf 1'''} = {\bf 1'},~~
{\bf 1''} \times {\bf 1'''}= {\bf 1},\nn\\
{\bf 1'} \times {\bf 1'''} &=& {\bf 1''},~~{\bf 1'} \times {\bf 1''} = {\bf 1'''},~~
{\bf 2} \times {\bf 1'} = {\bf 2},~~~~~ {\bf 2} \times {\bf 1''}
= {\bf 2'},\label{multi1}\\
{\bf 2} \times {\bf 1'''} &= &{\bf 2'}, ~~
{\bf 2'} \times {\bf 1'} = {\bf 2'}, ~~~~{\bf 2'} \times {\bf
1''} = {\bf 2},~~{\bf 2'} \times {\bf 1'''} ={\bf 2}\nn~.
\end{eqnarray}
Note that ${\bf 1''},~{\bf 1'''}$ and ${\bf 2}$ are
complex--valued.  The complex conjugate representation ${\bf 2^*}$
transforms the same way as ${\bf 2}$, when ${\bf 2^*}$ is
identified as ${\bf 2^*} = i \tau_2 {\bf 2}$, with $\tau_2$ being
the second Pauli matrix.
The Clebsch-Gordan coefficients for multiplying any of the  irreps
(which can be straightforwardly computed from the two-dimensional
representation of $A_{Q_N}$ and $B_Q$ given in (\ref{AD}) and (\ref{BD}))
are given by
\be
 \begin{array}{ccccccccc}
 {\bf 2}  &  \times   
&  {\bf 2}  &  =  &  {\bf 1} 
&  +   &  {\bf 1'} & + &  {\bf 2'} 
\\ 
 \left(\begin{array}{c} x_1 \\ x_2  \end{array} \right)   & 
 \times    &  \left(\begin{array}{c} y_1 \\  y_2  \end{array} \right)  
&  =  &   (x_1 y_2 - x_2 y_1)   &  &
 (x_1 y_1 +x_2 y_2)   &    &
 \left(\begin{array}{c} -x_1 y_2 - x_2 y_1 \\ x_1 y_1 - x_2 y_2  \end{array} \right) ,\\ 
\end{array}
  \label{multi2}
\ee
\be
\begin{array}{ccccccccc}
 {\bf 2'}  &  \times   
&  {\bf 2'}  &  =  &  {\bf 1} 
&  +   &  {\bf 1'} & + &  {\bf 2'} 
\\ 
 \left(\begin{array}{c}a_1 \\ a_2  \end{array} \right)   & 
 \times    &  \left(\begin{array}{c}b_1 \\  b_2   \end{array}\right)  
&  =  &   (a_1 b_1 + a_2 b_2)   &  &
 (a_1 b_2 -a_2 b_1)   &    &
 \left(\begin{array}{c}-a_1 b_1 + a_2 b_2 \\ a_1 b_2 +a_2 b_1 \end{array} \right) ,\\ 
\end{array}
\label{multi3} 
\ee
\be
\begin{array}{ccccccccc}
 {\bf 2}  &  \times   
&  {\bf 2'}  &  =  &  {\bf 1''} 
&  +   &  {\bf 1'''} & + &  {\bf 2} 
\\ 
 \left(\begin{array}{c}x_1 \\ x_2  \end{array} \right)   & 
 \times    &  \left(\begin{array}{c}a_1 \\  a_2 \end{array} \right)  
&  =  &   (x_1 a_2 + x_2 a_1)   &  &
 (x_1 a_1 -x_2 a_2)   &    &
 \left(\begin{array}{c}x_1 a_1 + x_2 a_2 \\ x_1 a_2 -x_2 a_1 \end{array} \right) .  \\ 
\end{array}
  \label{multi4}
\ee
In what follows we construct
a concrete model 
with the group theory rules given above.
Multiplication rules for $D_{3,4,6,8}, Q_{4,8}$are given
in Appendix.

\section{Supersymmetric extension of the standard model 
based on $Q_6$}
We would like to derive the mass matrix (\ref{nni}) 
solely from a symmetry principle.
 On finds that two conditions  should be met,
 as discussed already:
(i) There should be real as well as pseudo-real 
nonsinglet representations,
and  (ii)  there should be  the up- and down-type
 Higgs $SU(2)_L$ doublets (type II Higgs).
The smallest finite group that allows both real  and pseudo-real 
nonsinglet representations, is $Q_6$, as already found out.
One finds that all  Higgs fields can not be in real representations
to obtain the mass matrix (\ref{nni}).
So, certain Higgs fields are in pseudo-real representations. 
If  a Higgs field $H$ in a pseudo-real representation  
couples to a down-type  fermion family,
then its conjugate $H^*$ couples to a up-type fermion family.
Since $H^*$
belongs to  the conjugate representation, 
the mass matrix of the up-type family differs from that of
the down-type family. Therefore,  $H^*$ 
should be  forbidden to couple to the  up-type family
in order to obtain the same form of the mass matrices
for  the up and down sectors.
Accordingly, the Higgs sector
of the MSSM  fits  the desired Higgs structure.
Therefore, we concentrate on $Q_6$ and 
assume $N=1$ supersymmetry
 in the following discussions.

\subsection{$Q_6$ assignment}
Let us denote the fermion and Higgs fields as
\begin{eqnarray}
\psi &=& \left( \begin{array}{c}
\psi_1 \\ \psi_2 \end{array}\right),~~~ \psi^c =
\left(\begin{array}{c}-\psi_1^c \\ \psi_2^c \end{array}\right), ~~~\psi_3,~~~\psi_3^c,
\nonumber \\
H &=& \left(\begin{array}{c} H_1 \\ H_2 \end{array}\right),~~~H_3,
\label{def1}
\end{eqnarray}
and assume that  the assignment under ${\boldmath Q_6}$ is given by
\begin{eqnarray}
\psi: {\bf 2},~~~\psi^c: {\bf 2'},~~~\psi_3: {\bf 1'},~~~\psi_3^c:
{\bf 1'''}, ~~~H: {\bf 2'},~~~H_3: {\bf 1'''}
\label{assignment}
\end{eqnarray}
for all sectors.  Here $H^d$ and $H_3^d$ will couple to the down
quarks and charged leptons, while $H^u$ and $H_3^u$ will couple to
the up-quark and neutrino Dirac sectors.
The assignment under ${\boldmath Q_6}$ is the same in all sectors.
So we focus on the down-quark sector. 
From the  Clebsch-Gordan factors given in (\ref{multi1})
- (\ref{multi4}), the following Yukawa couplings
are found to be invariant:
\begin{equation}
{\cal L}_Y = a \psi_3\psi_3^c H_3 + b \psi^T \tau_1 \psi^c_3 H -
b' \psi_3  \psi^{cT} i\tau_2 H  + c \psi^T \tau_1 \psi^c H_3 + h.c.
\label{Ly}
\end{equation}
This leads to the mass matrix
\begin{eqnarray}
M=\left(\begin{array}{ccc} 0 & c\left\langle H_3\right\rangle  & b
\left\langle H_2\right\rangle \\ -c \left\langle H_3\right\rangle
& 0 & b \left\langle H_1\right\rangle \\  b'\left\langle
H_2\right\rangle & b'\left\langle H_1\right\rangle & a
\left\langle H_3\right\rangle \end{array}\right)~.
\label{nni2}
\end{eqnarray}
If $\left\langle H_2\right\rangle $ vanishes, the above mass matrix has
exactly the desired form (\ref{nni}).
If $\left\langle H_1\right\rangle 
=\left\langle H_2\right\rangle $,  we can bring 
the mass matrix (\ref{nni2}) to the desired form (\ref{nni}) by  
 an overall $45^0$ rotation on the fields
of the $Q_6$ doublets fermions (the $\psi$ and the $\psi^c$ fields).
So,
the mass matrix (\ref{nni2}) does not automatically lead to 
the desired form (\ref{nni});
 we need to construct
 a Higgs sector that ensures
 the stability of the desired VEV structure  
 $<H_1>=<H_2>$ (or $<H_2>=0$).
 A parity invariance $H_1 \leftrightarrow H_2$ would
 ensure the VEV structure, but it is not a symmetry of
 (\ref{Ly}). Therefore, the parity invariance can only be an
 accidental symmetry of the Higgs sector.
In section III. E we will construct a Higgs sector
that  possesses accidentally the desired parity invariance, 
while the whole sector is
invariant under $Q_6 \times  Z_{12R}$.
It would not work if the $Q_6$ doublets are replaced by doublets
of $D_{3,4,5,6,7}, Q_4$ or  $SU(2)$.

In Table 1 we write the $Q_6$ assignment (\ref{assignment}) again
to fix our notation, where $Z_{12R}$ will be introduced later on when constructing
the Higgs sector.
\vspace{0.5cm}
\begin{center}
\begin{tabular}{|c|c|c|c|c|c|c|}
\hline
 & $Q,L$ 
& $U^c,D^c,E^c,N^c$ & $H^u,H^d$ & $Q_3,L_3$  
& $U^c_3,D^c_3,E^c_3,N_3$  
& $H^u_3,H^d_3$  
\\ \hline
$Q_6$ &${\bf 2}$ & ${\bf 2'}$ &${\bf 2'}$ &
${\bf 1'}$  & ${\bf 1'''}$ &
${\bf 1'''}$ 
\\ \hline
$Z_{12R}$ & $1$ 
& $-1$ & $-2$ & $1$  & $-1$  
& $-2$  
\\ \hline
\end{tabular}
\vspace*{1mm}

{\bf Table 1}. $Q_6 \times Z_{12R}$ assignment 
of the matter supermultiplets. $Z_{12R}$ will be introduced to construct
a desired Higgs sector later on.
\end{center}

\subsection{Spontaneous CP violation}
The most stringent constraints on the soft supersymmetry breaking 
(SSB) parameters come from the electric dipole moments (EDM)
of neutron, electron and mercury atom.
Recent experiments on EDMs require that
$d_e < 64.3\times 10^{-27}$ e cm \cite{electron},
$d_n < 6.3\times 10^{-26}$ e cm \cite{neutron} and
$d_{H_g} < 2.1\times 10^{-28}$ e cm \cite{hg}.
A misalignment of the phases appearing in the Yukawa 
and  A terms (trilinear couplings of 
bosonic superpartners) is 
the origin of EDMs and  lead to a large contribution to EDMs, so that
a very fine tuning to suppress the misalignment is required \cite{gabbiani1,fcnc-edm,abel1,babu5}.
Note, however,  that a flavor symmetry can not ensure the alignment
of the phases.
Therefore,  we require that CP is  spontaneously broken
so that the origin of the CP phases
in the Yukawa  and  A terms is the same.
Consequently, all the coefficients appearing in the Lagrangian
should be  real.  (Some of
the couplings may be purely imaginary if the relevant fields
are odd under CP.) 
Then tanks to $Q_6$ family symmetry 
the phases 
in the Yukawa and  A terms after spontaneous
symmetry breaking of CP are so aligned,  that 
the contribution to EDMs coming from
the A terms in this model vanishes.
In the last section, we will
discuss the phase alignment again when discussing the 
SUSY flavor problem.
In the following discussions we simply assume that
\be
\left\langle H_1^u \right\rangle &=&\left\langle H_2^u \right\rangle,~
\left\langle H_1^d \right\rangle=\left\langle H_2^d \right\rangle,~
\left\langle H_3^u \right\rangle \neq 0,~
\left\langle H_3^d \right\rangle \neq 0,
\label{vev1}
\ee
and
denote the phases of VEVs as
\be
\Delta \theta^u &=&\arg \left\langle H^u_3 \right\rangle
-\arg \left\langle H^u_1 \right\rangle, \nn\\
\Delta \theta^d &=& \arg \left\langle H^d_3 \right\rangle
-\arg \left\langle H^d_1 \right\rangle.
\label{arg1}
\ee

In section III. E, we will construct a Higgs sector
 that satisfies  the assumptions (\ref{vev1}) made on VEVs. 
The crucial point is that the superpotential (\ref{WHp}) posses  the symmetry
(\ref{parity}), which can ensure the desired VEV structure (\ref{vev1}).
Note that the parity invariance (\ref{parity}) is not imposed by hand
on the superpotential (\ref{WHp}); 
it is an accidentally symmetry in the Higgs sector,
while the whole theory possesses 
$Q_6 \times  Z_{12R}$ symmetry only.

\subsection{Quark sector}
$Q_6$ assignments of  the left-handed quark supermutiplets
$Q$ and $Q_3$, and
the right-handed quark supermutiplets $U_R^c, U_{3R}^c, D_R^c, D_{3R}^c$
are given in Table 1. The most general $Q_6$ invariant 
superpotential for the Yukawa interactions in the quarks sector
is  $W_Q = W_D + W_U$, where
\be
W_D &=&
Y_a^d Q_3 D_{3}^c H^d_3 + Y_b^d Q^T \tau_1 D_{3}^c  H^d
 - Y^d_{b'} Q_3  D^{cT} i \tau_2  H^d 
+ Y^d_{c} Q^T  \tau_1 D^c H^d_3,
\label{potd}
\ee
and similarly for $W_U$ ( all the couplings are real).
Note that neither $Z_{12R}$ nor $R$ parity is assumed
to obtain $W_Q$. In fact,  including the
lepton sector, $Q_6$ alone forbids
all the $R$ parity violating  renormalizable terms, except one term.
It  also forbids  all the Baryon number violating Yukawa couplings.
We will come to discuss this in the last section.

We assume that VEVs take the form (\ref{vev1}), from which
we obtain the mass matrix for the up and down quarks.
We make   an overall $45^0$ rotation  on
 the $Q_6$ doublets, $Q, D^c$ and $U^c$, and 
phase rotations on the fields defined as
\be
U & \to & P_{u} U, U^c \to P_{u^c} U^c, 
\ee
and similarly for $D$ and $D^c$,
where
\be
P_{u,d} &=&    \left( \begin{array}{ccc}
1 & 0 & 0\\0 & \exp (i2\Delta \theta^{u,d}) & 0\\
 0 & 0&\exp (i\Delta \theta^{u,d})
\end{array}\right),
\label{Pu}\\
P_{u^c,d^c} &=&
\left( \begin{array}{ccc}
\exp (-i2\Delta \theta^{u,d})  & 0 & 0\\0 & 1& 0\\
 0 & 0&\exp (-i\Delta \theta^{u,d})
\end{array}\right)  
\exp \left( -i \arg \left\langle H^{u,d}_3\right\rangle \right).
\label{Puc}
\ee
($\Delta \theta^{u,d}$ are given in (\ref{arg1}).)
Then one obtains  real mass matrices
\be
M_{u,d} &=& m_{t,b}
 \left( \begin{array}{ccc}
0 & q_{u,d}/y_{u,d} & 0\\- q_{u,d}/y_{u,d} & 0 & b_{u,d}\\
 0 & b_{u,d}' & y_{u,d}^2
\end{array}\right),
\label{Mud}
  \ee
where  $m_t$ and $m_b$ are the top and bottom quark mass, 
respectively \footnote{
Ma \cite{ma0} considered a model based on a $S_3 \times  Z_3$
family symmetry. In his model, the mass matrix for
the down quarks is similar to $M_d$, but that for the
up quark sector is diagonal. So tow models are distinguishable.}
We have changed our notation for the mass matrices
to directly apply the result of \cite{harayama1} to our case.
The CKM matrix $V_{\rm CKM}$ is
  given by 
 \be
 V_{\rm CKM} &=& O^T_u P_q O_d,
 \label{vckm}
 \ee
 where
 \be
 O^T_u M_u M_u^T O_u &=&
  \left( \begin{array}{ccc}
m_u^2 & 0 & 0\\0 & m_c^2 & 0\\
 0 & 0&m_t^2
\end{array}\right),~
 O^T_d M_d M_d^T O_d =
  \left( \begin{array}{ccc}
m_d^2 & 0 & 0\\0 & m_s^2 & 0\\
 0 & 0&m_b^2
\end{array}\right),
\label{Oud}\\
P_q &=& P_u^\dag P_d=
  \left( \begin{array}{ccc}
1 & 0 & 0\\0 & \exp (i2\theta_q) & 0\\
 0 & 0&\exp (i\theta_q)
\end{array}\right)~\mbox{with}~\theta_q=\Delta \theta^d-\Delta \theta^u.
\label{Pq}
\ee
Note that there are only 8 real independent parameters,
i.e., $q_{u,d}, b_{u,d}, b_{u,d}', y_{u,d}$, and one
independent phase $\theta_q$ to describe $V_{\rm CKM}$ and the quark masses.
Using the result of \cite{harayama1} we find that $V_{\rm CKM}$ 
can be approximately written as
 \be
  V_{ud} &\simeq &1,   V_{cs} \simeq \exp i 2\theta_q,   
  V_{tb} \simeq \exp i \theta_q, 
  \label{vud}\\
 V_{us} &\simeq & -y_d \sqrt{\frac{m_d}{m_s}}+y_u\sqrt{\frac{m_u}{m_c}}\exp i 2 \theta_q,
 \label{vus}\\
 V_{cb} &\simeq & \frac{y_d^2}{\sqrt{1-y_d^4}} \frac{m_s}{m_b}\exp i 2 \theta_q
 - \frac{y_u^2}{\sqrt{1-y_u^4}}\frac{m_c}{m_t}\exp i  \theta_q,
  \label{vcb}\\
 V_{ts} &\simeq & -\frac{y_d^2}{\sqrt{1-y_d^4}} \frac{m_s}{m_b}\exp i  \theta_q
 + \frac{y_u^2}{\sqrt{1-y_u^4}}\frac{m_c}{m_t}\exp i  2\theta_q,
  \label{vts}\\
 V_{ub} &\simeq & \frac{\sqrt{1-y_d^4}}{y_d}\sqrt{\frac{m_d}{m_s}}\frac{m_s}{m_b}\nn \\
 & &+
y_u \sqrt{\frac{m_u}{m_c}}\left(
\frac{y_d^2}{\sqrt{1-y_d^4}} \frac{m_s}{m_b}\exp i 2 \theta_q
 - \frac{1}{y_u^2\sqrt{1-y_u^4}}\frac{m_c}{m_t}\exp i  \theta_q\right),
  \label{vub}
 \\
V_{td} &\simeq & \frac{\sqrt{1-y_u^4}}{y_u}\sqrt{\frac{m_u}{m_c}}\frac{m_c}{m_t}\nn\\
 & &+
y_d \sqrt{\frac{m_d}{m_s}}\left(
\frac{y_u^2}{\sqrt{1-y_u^4}} \frac{m_c}{m_t}\exp i 2 \theta_q
 - \frac{1}{y_d^2\sqrt{1-y_d^4}}\frac{m_s}{m_b}\exp i  \theta_q\right). 
  \label{vtd}
 \ee

For the mass matrix parameters (defined in (\ref{Mud}))
 \be
\theta_q & = &-1.15,
q_u=0.0002260,
b_u=0.04596,
b'_u=0.08959,
y_u=0.99746\label{thetaq},\\
q_d &= &0.005110,
b_d=0.02609,
b'_d=0.7682,
y_d=0.8000,\nn
\ee
we obtain
\be
m_u/m_t &=& 1.24\times 10^{-5},
m_c/m_t=4.13 \times 10^{-3},
m_d/m_b=1.22 \times 10^{-3},
m_s/m_b=2.14 \times 10^{-2},
\label{q-mass}\\
|V_{ud}| & = & 0.975 (1.00), |V_{us}| = 0.222 (0.231),
|V_{ub}|=3.64 (3.59)\times 10^{-3},
|V_{cb}|=4.20 (3.73)\times 10^{-2},\nn\\
|V_{td}| &=& 8.82 (8.75)\times 10^{-3},
|V_{ts}| = 4.12 (3.73)\times 10^{-2}, |V_{tb}| = 0.999 (1.00),\nn\\
\bar{\eta} &=& 0.361 (0.411),
\sin 2\beta (\phi_1)=0.704 (0.740),
\ee
where $\bar{\eta}$ is one of the  Wolfenstein parameters. The values in the parentheses
are those obtained from the approximate
formulae (\ref{vud})-(\ref{vtd}).
The experimental values are \cite{pdg}:
\be
|V_{\rm CKM}^{\rm exp}| &=&
\left( \begin{array}{ccc}
0.9739~\mbox{to}~ 0.9751 & 0.221 ~\mbox{to}~ 0.227 & 0.0029 ~\mbox{to}~ 0.0045
\\  0.221 ~\mbox{to}~ 0.227  & 0.9730 ~\mbox{to}~ 0.9744   & 0.039 ~\mbox{to}~ 0.044
  \\ 0.0048 ~\mbox{to}~ 0.014  &0.037 ~\mbox{to}~ 0.043  & 0.9990 ~\mbox{to}~ 0.9992
\end{array}\right),
\ee
and
 \be
\sin 2\beta (\phi_1)&=&0.736\pm0.049, \bar{\eta }=0.33 \pm 0.05.
 \ee
 The quark masses at $M_Z$ are given by \cite{kim1}
\be
m_u/m_d &=& 0.553\pm 0.086~(0.61)~,
~m_s/m_d = 18.9\pm 1.6~(17.5),\nn\\
~m_c  &=&0.73\pm 0.17~(0.72)~\mbox{GeV}~,
~m_s =0.058\pm 0.015~(0.062)~\mbox{GeV},\nn\\
~m_t &=& 175 \pm 6 ~\mbox{GeV}~,
~m_b = 2.91\pm 0.07~\mbox{GeV},
\label{qmass}
\ee
where the values in the parentheses are the theoretical values obtained
from  (\ref{q-mass}) for $m_t= 174  $ GeV and $m_b=2.9  $ GeV.
So, we see that the 9 independent parameters can well describe 10
physical observables. 
Our numerical analysis show that
the experimental  constraints
can be satisfied if
\be
-1.3 &< & \theta_q < -1.0~\mbox{and}~m_s/m_d < 19.
\label{range}
\ee

\begin{center}
\begin{figure}[htb]
\includegraphics*[width=0.6\textwidth]{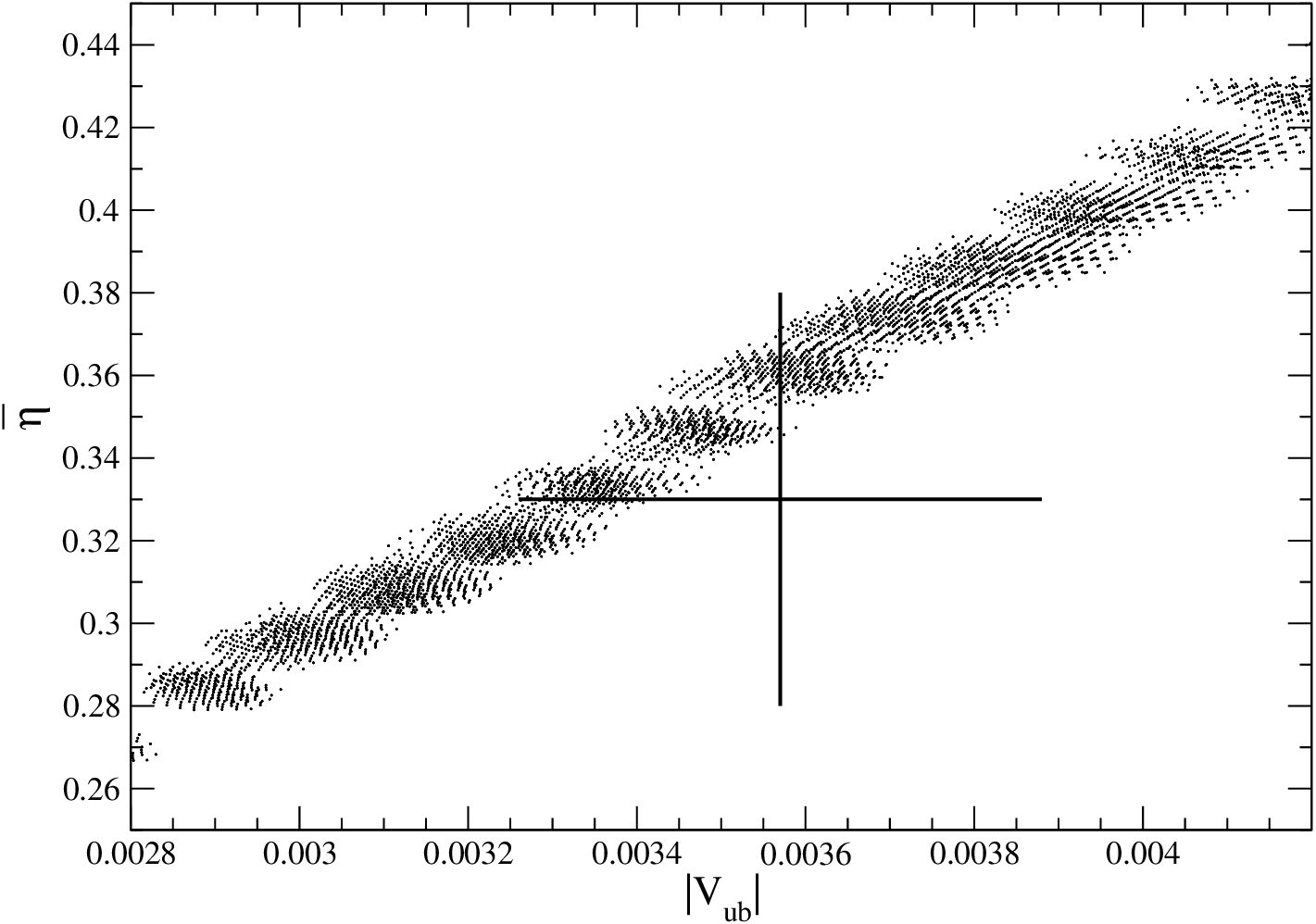}
\caption{\label{fig1}
Predictions in the $|V_{ub}|-\bar{\eta}$ plane.
The uncertainties result from those in the quark masses (\ref{qmass})
and in  $|V_{us}|$ and $|V_{cb}|$, where
we have used
$|V_{us}|=0.2240\pm 0.0036$ and $|V_{cb}|=(41.5\pm 0.8)\times 10^{-3}$ \cite{kim1}.
The vertical and horizontal lines correspond to
the experimental values 
$\bar{\eta}=0.33\pm 0.05$ and  $|V_{ub}|=(35.7\pm 3.1)\times 10^{-4}$ \cite{pdg,kim1}.}
\end{figure}
\end{center}

\begin{center}
\begin{figure}[htb]
\includegraphics*[width=0.6\textwidth]{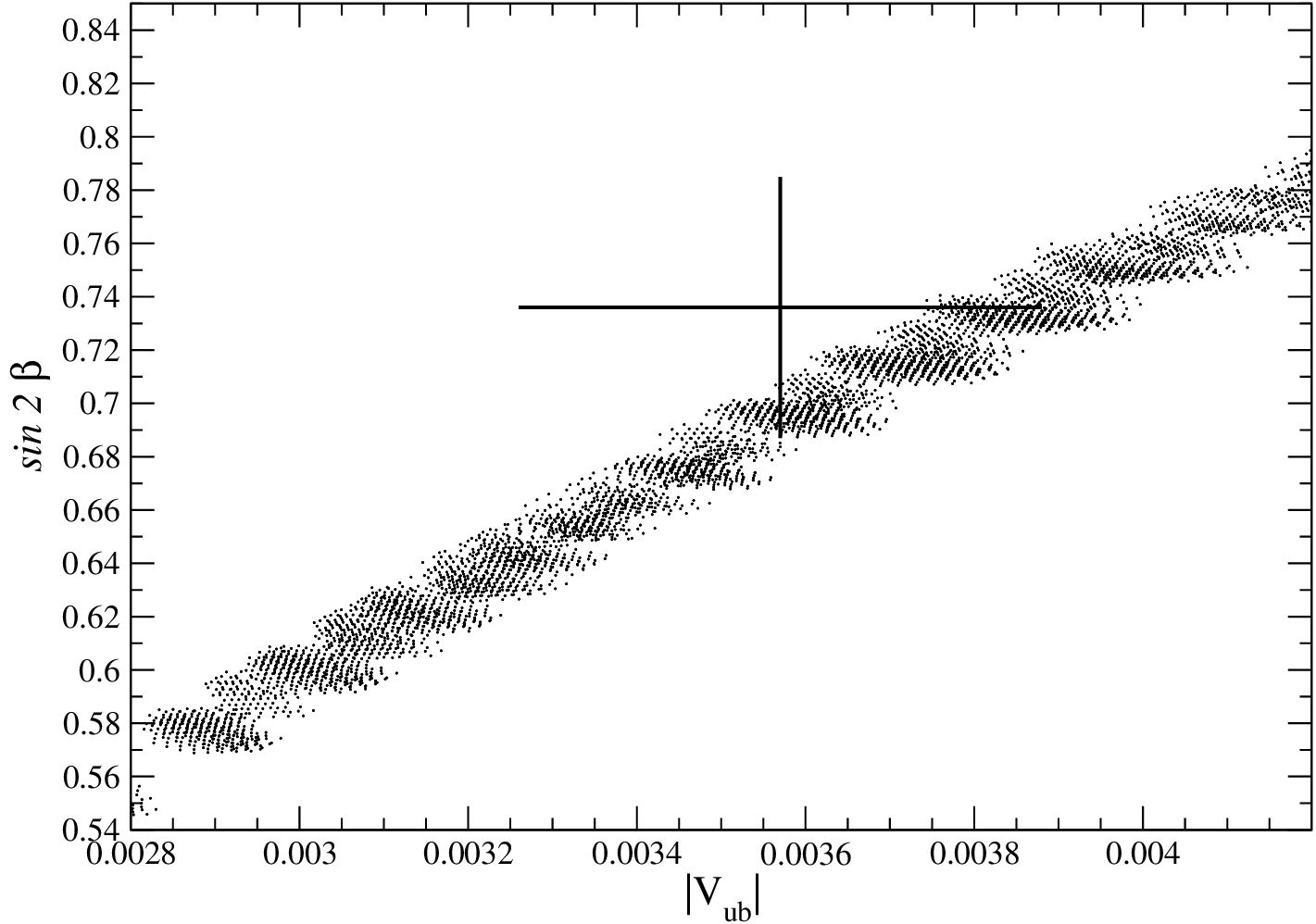}
\caption{\label{fig2}
Predictions in the $|V_{ub}|-\sin2\beta(\phi_1)$ plane.}
\end{figure}
\end{center}

In Fig.~1 we plot  predicted points
in the $|V_{ub}|-\bar{\eta}$ plane, where we have used
the quark masses  with the uncertainties given in (\ref{qmass})
and $|V_{us}|=0.2240\pm 0.0036,   
|V_{cb}|=(41.5\pm 0.8)\times 10^{-3}$ \cite{kim1}.
The experimental values of  $|V_{ub}|$ and $\bar{\eta}$
are indicated by the solid lines.
As we see from Fig.~1 and 2, accurate determinations of
$\bar{\eta},\sin2\beta(\phi_1)$ and $|V_{ub}|$ are crucial to test the quark sector of the model.

 In Fig.~3 we plot $|V_{td}/V_{ts}|$ against
$|V_{ub}|$.
From  the figure we find that
 \be
 \left| \frac{V_{td}}{V_{ts}} \right| & = &  0.205-0.230.
\label{vtdts}
\ee
The present experimental upper 
 bound is  $0.25$ \cite{pdg}, which is obtained from the ratio
 $\Delta M_{B_s}/\Delta M_{B_d}$, where 
  $\Delta M_{B_d}$ is well measured, but only 
  an lower bound on  $\Delta M_{B_s}$
  is experimentally known. 
 Note that a direct comparison of
 the theoretical values given in (\ref{vtdts}) 
 with the experimental value is possible, only if there is no
 contribution to   $\Delta M_{B_s}$ and $\Delta M_{B_d}$
 other  than those of the SM. In  supersymmetric extensions of the SM
 and in multi-Higgs models there are extra contributions in general.
 Nevertheless, $|V_{td}/V_{ts}|$ will be an important quantity
 for the $Q_6$ model.
 
\begin{center}
\begin{figure}[htb]
\includegraphics*[width=0.6\textwidth]{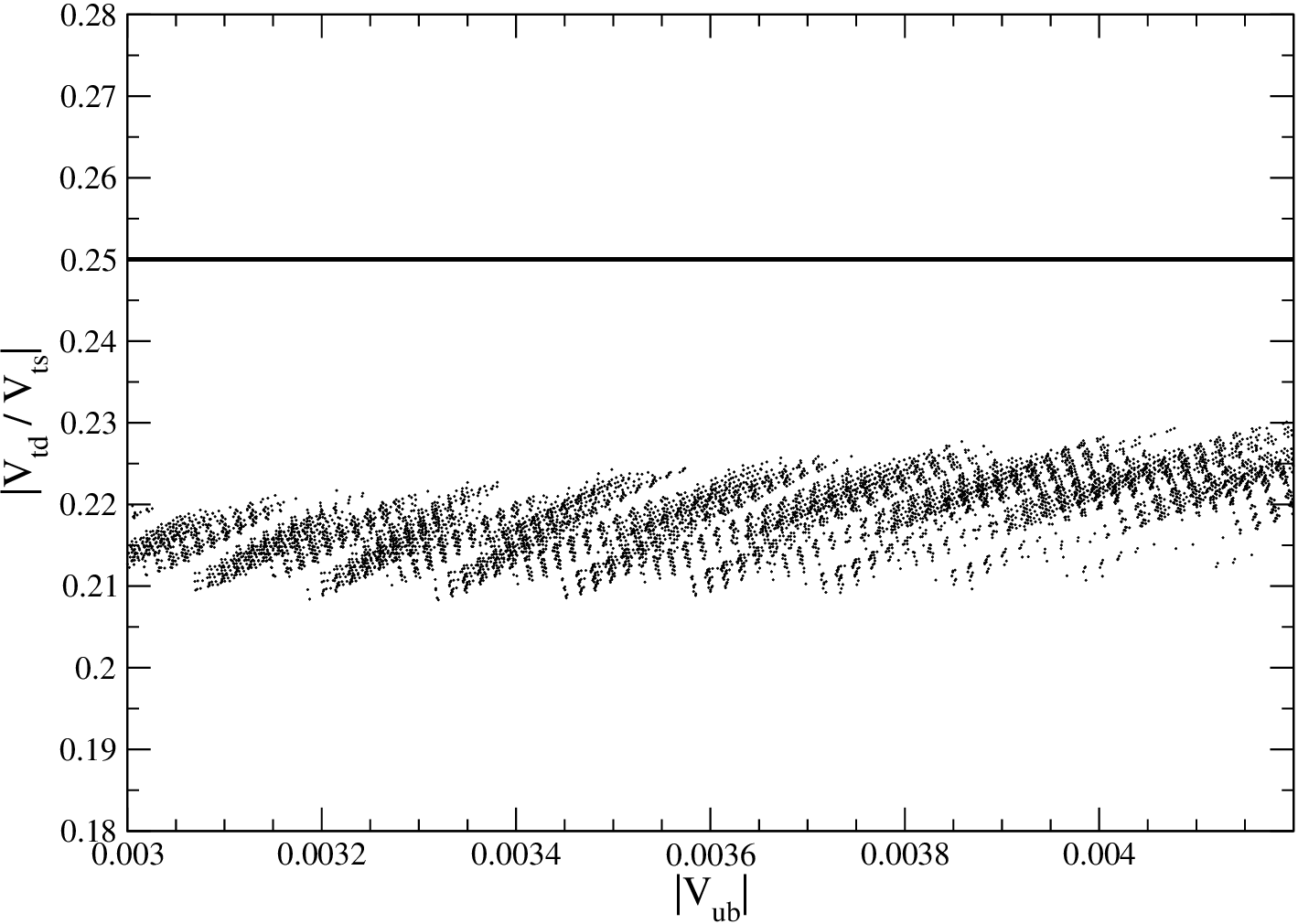}
\caption{\label{fig3}
Predictions in the $|V_{ub}|-|V_{td}/V_{ts}|$ plane.
The horizontal line is  the experiment upper bound.}
\end{figure}
\end{center}

\subsection{Lepton sector}
As in the quark sector, the $Q_6$ invariant 
superpotential for the Yukawa interactions in the lepton  sector
is given by $W_L = W_N + W_E$, where
\be
W_N &=&
Y_a^\nu L_3 N_{3}^c H^u_3 + Y_b^\nu L^T \tau_1 N_{3}^c  H^u
 - Y^\nu_{b'} L_3  N^{cT} i \tau_2  H^u
+ Y^\nu_{c} L^T\tau_1 N^c H^u_3,
\label{potn}
\ee
and similarly for $W_E$ \footnote{An alterative assignment of the leptons
is
\begin{eqnarray}
L: {\bf 2},~~E^c: {\bf 2},~~L_3: {\bf 1''},~~E_3^c:
{\bf 1''}, ~~N^c: {\bf 2}, ~~N_3^c:{\bf 1},\nn
\end{eqnarray}
where the  assignment of the quarks and Higgs supermultiplets 
remains unchanged.
The corresponding  Yukawa  sector 
will coincide with that of the $S_3$ model with $Z_2$
in the lepton sector \cite{kubo1,kubo2}. It contains only 6 real parameters with two phases
in the lepton sector and explains the maximal mixing of the atmospheric neutrinos.
In that model, there exists not only
Majorana phases but also  a Dirac-type  CP phase,
which was overlooked in \cite{kubo1,kubo2}.
Because of the spontaneous CP violation in the present case,
there is only one independent phase in the lepton sector.
So, the Dirac and Majorana phases are related.}.
All the couplings are real, because we require that
CP is only spontaneously broken.
The charged lepton sector is very similar to the quark sector, and 
we find that the diagonalizing unitary matrix $U_e$ 
rotating the left-handed fermions is given by
\be
U_e^\dag &=& O_e^T P_e^\dag,
\ee
where the orthogonal matrix $O_e$ can be  approximately
written as 
\be
O^T_e
& \simeq  & \left( \begin{array}{ccc}
1 & \sqrt{\frac{m_e\cos\phi }{m_\mu}}& - \sqrt{\frac{m_e }{ m_\mu }}\frac{\sin\phi}{\sqrt{\cos\phi}}
 \\ -  \sqrt{\frac{m_e }{m_\mu \cos\phi }} &   \cos\phi ~(1+
\frac{m_\mu^2}{m_\tau^2} )&
-\sin\phi ~(1- \frac{m_\mu^2}{m_\tau^2 \tan^2\phi} )  \\
\sqrt{\frac{m_e}{m_\mu }} \frac{m_\mu^2}{ m_\tau^2}\frac{\sqrt{\cos\phi}}{\sin\phi}
 & \sin\phi ~(1- \frac{m_\mu^2}{m_\tau^2 \tan^2\phi} )  & 
  \cos\phi ~(1+
\frac{m_\mu^2}{m_\tau^2} )
\end{array}\right),
\label{ote}
\ee
and
\be
P_e &=&
 \left( \begin{array}{ccc}
 1 & 0 &\\
 0 & \exp (i2 \Delta \theta^d) & 0\\
 0 & 0& \exp (i \Delta \theta^d)
 \end{array}\right).
 \label{Pe}
\ee
The angel $\phi$ is an independent free parameter, and it can be assumed
without loss of generality that
$\sin\phi$ and $\cos\phi$ are positive.

Now let us come to the neutrino sector.
Since $N^c$ belongs to ${\bf 1'''}$ of $Q_6$,
$N_3^c$ can not form a $Q_6$ invariant mass term, 
while the $Q_6$ doublet $N^c$ can do ($N_1^c N_1^c+N_2^cN_2^c$).
The absence of a mass term for $N_3^c$ would be phenomenologically
inconsistent.
In section III.E, when discussing the Higgs sector,
we will  give a way  to obtain a mass term
for $N_3^c$. Here we simply assume that that there is mass term for $N_3^c$.
So, the mass matrix $M_R$ of the right-handed neutrinos is 
diagonal: $M_R=\mbox{diag}(M_1,M_1,\kappa M_3)$,
where $\kappa=\pm 1$ takes into account the difference 
of the $Q_6$ doublet $N^c$ and  $Q_6$ singlet $N^c_3$ under CP.
After  an overall $45^0$ rotation on
 the $Q_6$ doublets, and 
using the sea-saw mechanism, we obtain a neutrino mass matrix.
Next we make a phase rotation 
on the left-handed neutrinos with
\be
P_\nu &=& 
 \left( \begin{array}{ccc}
 \exp (-i r_1)& 0 &\\
 0 &  \exp (-i r_2) & 0\\
 0 & 0&  \exp (-i r_3)
 \end{array}\right),\\
 \label{Pnu}
r_1 &=& r_2=\arg <H_3^u>, r_3=\arg <H_1^u>
\ee
to obtain the left-handed neutrino mass matrix
\be
M_{\nu} 
&=& 
 \left( \begin{array}{ccc}
c^2_\nu &  0  & c_\nu b_\nu' \\
0  & c^2_\nu+\kappa b^2_\nu  e^{- i  \varphi }& \kappa   a_\nu b_\nu  \\
 c_\nu b_\nu' & \kappa   a_\nu b_\nu   & \kappa a^2_\nu e^{ i \varphi } +b'^2_\nu
\end{array}\right),
\label{numass}\\
\varphi  &= & 2( \arg <H_3^u>-\arg <H_1^u>)=2 \Delta \theta^u.
\label{varphi}
\ee
\vspace{0.5cm}
\noindent
{\bf I. Inverted spectrum}

\vspace{0.5cm}
First we discuss the case of an inverted spectrum
of neutrino masses, i.e. $ m_{\nu_2} > m_{\nu_1} > m_{\nu_3}$.
To this end,  we assume that $b_\nu$ is small, and we treat it as perturbation:
\be
M_{\nu}
&=& M_{\nu}^{(0)}+  M_{\nu}^{(1)}+O(b_\nu^2)
\ee
where
\be
M_{\nu}^{(0)}&=& \left( \begin{array}{ccc}
c^2_\nu &  0  & c_\nu b_\nu' \\
0  & c^2_\nu  &0  \\
 c_\nu b_\nu' & 0   & \kappa a^2_\nu e^{ i \varphi } +b'^2_\nu
\end{array}\right),
M_{\nu}^{(1)}= \left( \begin{array}{ccc}
0 &  0  & 0 \\
0  & 0 & \kappa a_\nu b_\nu  \\
0  & \kappa a_\nu b_\nu  & 0
\end{array}\right).
\label{numass0}
\ee
Because the $(1,2)$ and $(2,1)$ elements of $M_{\nu}$ is exactly zero,
the correction to the eigenvalues is $O(b_\nu^2)$. But the correction to
the unitary matrix $U_\nu$ is $O(b_\nu)$. This correction 
is found to be 
\be
U_\nu &=& U_\nu^{(0)}+ U_\nu^{(1)}+O(b^2),
\label{unu1}\\
 U_\nu^{(1)} &=&
 \left( \begin{array}{ccc}
0 &  0  & \delta U_{13}\\
\delta U_{21}  &  \delta U_{22} &  0 \\
0  & 0  & 0
\end{array}\right),\\
 \delta U_{13} &=&  -\kappa  a_\nu b_\nu /c b'_\nu, \nn\\
 \delta U_{21} & = &-(\cos\theta_\nu)( e^{ -i (\varphi_1-\phi_\nu )/2})\delta U_{13},\nn\\
 \delta U_{22} & = &(\sin\theta_\nu)( e^{ -i (\varphi_2-\phi_\nu )/2})\delta U_{13},\nn
 \ee
 where $ U_\nu^{(0)}, \varphi_{1,2}, \phi_\nu$ and $\theta_\nu$ are given below.
The unitary matrix $ U_\nu^{(0)}$ can be obtained from
\be
U^{(0)T}_\nu M_{\nu}^{(0)} U_\nu^{(0)} &=& \left( \begin{array}{ccc}
m_{\nu_1} & 0 & 0\\
0 & m_{\nu_2} &0 \\
0 & 0 & m_{\nu_3}
\end{array}\right),
\ee
where $m_{\nu_i}$ are real and positive, and 
\be
U_{\nu}^{(0)}&= &\left( \begin{array}{ccc}
c_{\nu}e^{-i (\varphi_1-\phi_\nu)/2}  & -s_{\nu}e^{-i (\varphi_2-\phi_\nu)/2} 
&  0
\\ 0 & 0 &1
\\  s_{\nu}e^{-i (\varphi_1+\phi_\nu)/2}   &
 c_{\nu}e^{-i (\varphi_2+\phi_\nu)/2}   & 0
 \end{array}\right)
 \label{unumax3}
 \ee
with $c_\nu=\cos\theta_\nu$ etc.
Note that we obtain  an inverted spectrum
\be
m_{\nu_1},m_{\nu_2} > m_{\nu_3}
\label{spectrum}
 \ee
 as long as $b_\nu$ is small.
The phases and the masses satisfy to $O(b_\nu)$ the following relations:
 \be
m_{\nu_3} \sin \phi_\nu &=& m_{\nu_2} \sin \varphi_2
=m_{\nu_1} \sin \varphi_1,
\label{numasses}\\
\varphi &=&2\Delta \theta^u
 = \varphi_{1}+\varphi_{2} 
-\frac{1}{2}(1-\kappa) \pi +(\mbox{mod}~2\pi)\label{varphi2} \\
& \simeq &
\frac{1}{2}(1+\kappa) \pi +(\mbox{mod}~2\pi),\nn\\
m_{\nu_3} &=& c^2_\nu,~
\frac{m_{\nu_1}m_{\nu_2}}{m_{\nu_3}}= a^2_\nu,
\label{rho23}\\
\tan \phi_\nu &=&
-\frac{\kappa a^2_\nu \sin\varphi}{c^2_\nu+b'^2_\nu
+\kappa a^2_\nu\cos \varphi},
\label{tanvarphi}\\
\tan^2\theta_{\nu} &=&
\frac{(m_{\nu_1}^2-m_{\nu_3}^2 \sin^2\phi_\nu)^{1/2}
-m_{\nu_3}|\cos\phi_\nu|}{(m_{\nu_2}^2
-m_{\nu_3}^2 \sin^2\phi_\nu)^{1/2}
+m_{\nu_3}|\cos\phi_\nu|}.
\label{theta-m}
\ee
The last equation (\ref{theta-m}) can be rewritten as
\be
\frac{m_{\nu_1}^2}{\Delta m_{23}^2-\Delta m_{21}^2} &=&
\frac{(1+2 t_{\nu}^2+t_{\nu}^4+r t_{\nu}^4)^2}
{4  t_{\nu}^2 (1+t_{\nu}^2)(1+t_{\nu}^2+r t_{\nu}^2)\cos^2 \phi_\nu}
-\tan^2 \phi_\nu\nn\\
&\simeq &
\frac{1}{\sin^2 2\theta_{12}\cos^2 \phi_\nu}
-\tan^2 \phi_\nu ~~\mbox{for}~~|r| << 1,
\label{mnu21}
\ee
where $t_\nu=\tan\theta_\nu$ and $r=\Delta m_{21}^2/(\Delta m_{23}^2-\Delta m_{21}^2)$.

The mixing matrix is  obtained from
$O^T_e P_e^\dag  P_{\nu} U_\nu$, which can be brought into the form
\be
V_{\rm MNS}  &=& \hat{V}_{\rm MNS}
\times
\left( \begin{array}{ccc}
1 & 0 & 0\\
0 & e^{i \alpha} &0 \\
0 & 0 & e^{i \beta}
\end{array}\right),
\label{vmns}
\ee
where $(1,1),(1,2),(2,3)$ and $(3,3)$ elements 
of $\hat{V}_{\rm MNS}$ are real. $O^T_e,  P_e,  P_\nu$ and $ U_\nu$
are given in (\ref{ote}), (\ref{Pe}), (\ref{Pnu}) and (\ref{unu1}).
To $O(b_\nu)$, the mixing matrix $\hat{V}_{\rm MNS}$ and 
the Majorana phases are approximately given by
\be
\alpha &=&\frac{1}{2}(\varphi_1-\varphi_2)
\label{alpha}\\
\beta &=&-\Delta\theta^u-\Delta \theta^d+ \frac{1}{2}(\varphi_1+\phi_\nu)
\label{beta}\\
\hat{V}_{e2}
&\simeq & -\sin\theta_{12} \simeq -\sin \theta_\nu,\\
\hat{V}_{\mu3} 
&\simeq &\cos\theta_{23} \simeq \cos\phi(1+m_{\mu}^2/m_{\tau}^2),
\label{vmu3}\\
\hat{V}_{\tau3}
&\simeq  & \sin\theta_{23} \simeq\sin\phi(1-m_{\mu}^2/m_{\tau}^2\tan^2\phi),\\
\hat{V}_{e3}
&=  & -\sin\theta_{13}e^{-i\delta_{CP}}
\label{dcp}\\
&\simeq& -\kappa(a_\nu b_\nu/c_\nu b'_\nu) e^{-i\delta_{1CP}}+
(m_{e}\cos\phi/m_{\mu})^{1/2}e^{-i \delta_{2CP}},\\
\label{ve3}
\delta_{1CP}&=&\phi_\nu-\Delta\theta^u-\Delta \theta^d 
\label{d1},\\
\delta_{2CP} &=&\phi_\nu-\Delta\theta^u+\Delta \theta^d,
\label{d2}
\ee
where $\phi $ is defined in (\ref{ote}),
 and $\phi_\nu$ and $\theta_\nu$ are defined in  (\ref{unumax3}).
Since $\varphi=2\Delta\theta^u$ (see (\ref{varphi})) and because 
of (\ref{numasses}) and (\ref{tanvarphi}),
only $\Delta\theta^u$ and $\Delta \theta^d$ are independent phases.
Note that  $\varphi_2+\varphi_1 \simeq \pi$ and
$\varphi_2 \sim\pi$. If $\varphi_1 \sim\pi$, 
we obtain $\tan\theta_\nu > 1$.

Although $b_\nu$ is a free parameter, it is possible
to give a minimum for $|V_{e3}| $,
because the difference of the phases $\delta_{1CP}$ and $\delta_{2CP}$ 
depends  only on $\Delta \theta^d$:
\be
|V_{e3}| &=&
|\hat{V}_{e3}| \simeq |\sin \theta_{13}|
>  (m_{e}\cos\theta_{23}/m_{\mu})^{1/2}|\sin 2\Delta \theta^d|\\
&  >  &0.057 |\sin 2\Delta \theta^d| \sim 0.03,
\label{exp-v}
\ee
where we have used (\ref{vmu3}),  (\ref{d1})
and (\ref{d2}), $-1.3 < \theta_q=\Delta \theta^d-
\Delta \theta^u < -1.0$ (see (\ref{range})) and $2 \Delta \theta^u
\simeq  \pi~\mbox{or}~0$ (see (\ref{varphi2})).

\begin{center}
\begin{figure}[htb]
\includegraphics*[width=0.6\textwidth]{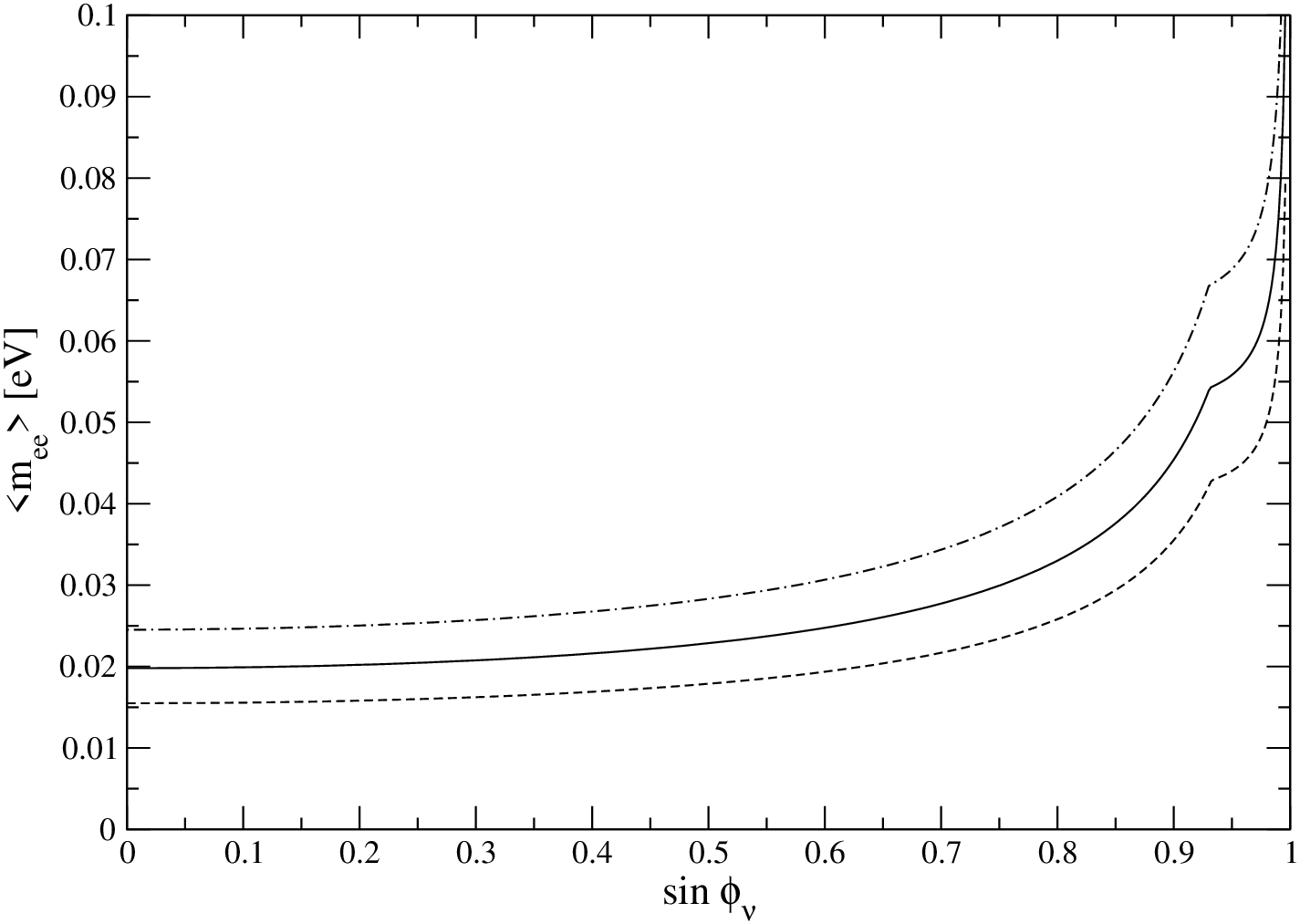}
\caption{\label{fig4}
 The average neutrino mass $<m_{ee}>$ in
 the limit $b_\nu \to 0$ as a function of 
 $\sin\phi_\nu$ with $\sin^2\theta_{12}=0.3$ and
 $\Delta m_{21}^2=8.1\times \mbox{eV}^2$.
  The dashed, solid and dot-dashed lines correspond to
 $\Delta m_{23}^2=1.4, 2.2$ and $ 3.3 \times 10^{-3}$ eV$^2$,
respectively. The $\Delta m_{21}^2$ dependence is very small.
The neutrino spectrum is inverted.}
\end{figure}
\end{center}

\begin{center}
\begin{figure}[htb]
\includegraphics*[width=0.6\textwidth]{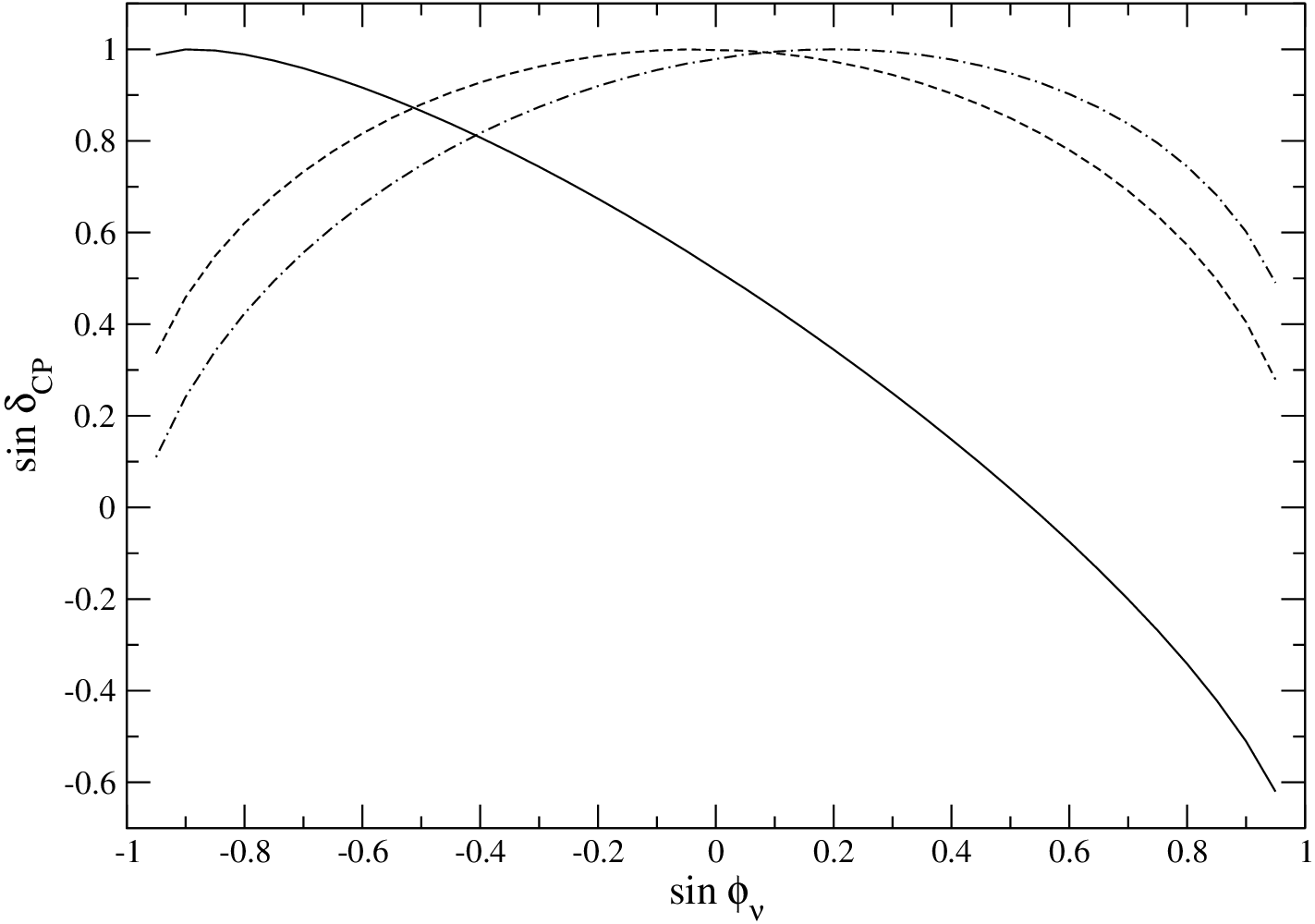}
\caption{\label{fig5}
$\sin\delta_{CP}$   in
 the limit $b_\nu \to 0$ versus $\sin\phi_\nu$
for different values of $\sin\theta_{13}$, where we use
$\sin^2\theta_{12}=0.3, \theta_q=-1.15$. 
The solid line corresponds to $\sin\theta_{13}=0.0438$,
which is the minimum for $\theta_q=-1.15$.
The other lines are those for $\sin\theta_{13}=
0.10$ (dotted), $0.20$ (dot-dashed), respectively. $\theta_q$ is defined in (\ref{numass}). 
The result does not depend on $\kappa$.}
\end{figure}
\end{center}

Using (\ref{numasses}) one can express to $O(b_\nu)$
$\sin 2 \alpha$  as
\be
\sin 2 \alpha &=&\sin(\varphi_1-\varphi_2)\nn\\
& =&
-\frac{ m_{\nu_3}\sin\phi_\nu}{m_{\nu_1}m_{\nu_2}}
\left( \sqrt{m_{\nu_2}^2-m_{\nu_3}^2 \sin^2 \phi_\nu}+
\sqrt{m_{\nu_1}^2-m_{\nu_3}^2 \sin^2 \phi_\nu} \right)
\label{alpha1},
\ee
where we have used $\varphi_2+\varphi_1 \simeq \pi$ and
$\varphi_2 \sim\pi$.
Eq. (\ref{tanvarphi}) can be rewritten in  terms of measurable quantities:
\be
\kappa\sin \varphi &=&\kappa\sin 2\Delta\theta^u=
\frac{m_{\nu_{3}}}{m_{\nu_{1}}} \sin\phi_\nu
\left[ 1-(\frac{m_{\nu_{3}}}{m_{\nu_{2}}}\sin\phi_\nu)^2\right]-
\frac{m_{\nu_{3}}}{m_{\nu_{2}}} \sin\phi_\nu
\left[ 1-(\frac{m_{\nu_{3}}}{m_{\nu_{1}}}\sin\phi_\nu)^2\right]\nn\\
&\simeq&
\frac{m_{\nu_{3}}}{2m_{\nu_{2}}^{3}}
\frac{\Delta m_{21}^{2}\sin\phi_{\nu}}{
(1-(m_{\nu_{3}}/m_{\nu_{2}})^{2}\sin^{2}\phi_{\nu})^{1/2}},
\label{varphi3}
\ee
which relates $\phi_{\nu}$ with $\Delta\theta^u$.
Although  $\sin \varphi \simeq 0$ because of (\ref{varphi2}),
$\sin\phi_\nu$ is not small and can vary from $-1$ to $+1$.
So, it may be better to regard $\sin\phi_\nu$ 
as independent instead of $\Delta\theta^u$.

The average neutrino mass $<m_{ee}>$
can be then predicted from
\be
<m_{ee}> &=& |~\sum_{i=1}^3m_{\nu_i} V_{ei}^2|
\simeq |m_{\nu_1}c_{\nu}^2+
m_{\nu_2}s_{\nu}^2 \exp i2\alpha~ |.
\label{mee}
\ee
To obtain numerical values of $<m_{ee}>$, we use the 
experimental values  \cite{maltoni1}
\be
7.2\times 10^{-5}  &\mbox{eV}^2 & \leq  \Delta m^2_{21} \leq 9.1 \times 10^{-5}  \mbox{eV}^2,
1.4\times 10^{-3}  \mbox{eV}^2  \leq \Delta m^2_{23} \leq  3.3 \times 10^{-3}  \mbox{eV}^2,\\
0.23 & \leq &
\sin^2 \theta_{12} \leq 0.38,~0.34  \leq  \sin^2 \theta_{23} \leq  0.68,~
|\sin^2 \theta_{13}| < 0.047.
\ee
Note that $\alpha$ is given in (\ref{alpha1}) and the absolute scale of the neutrino
mass can be obtained from (\ref{mnu21}).
In Fig.~4 we  plot $<m_{ee}>$  as a function of $\sin \phi_\nu$
for $\sin^2\theta_{12}=0.3, \Delta m_{21}^2=8.1 \times10^{-5}$ eV$^2$ and
$\Delta m_{23}^2=1.4, 2.2, 3.3 \times 10^{-3}$ eV$^2$.
As we can see from Fig.~2, the effective Majorana mass stays
at about its minimal value $<m_{ee}>_{\rm min}$ for a wide range of $\sin\phi_\nu$.
Since $<m_{ee}>_{\rm min}$ is 
approximately equal to
$<m_{ee}>_{\rm min} \simeq \sqrt{\Delta m^2_{23}}/\tan 2\theta_{12}
=   ( 0.010 - 0.037 ) ~~\mbox{eV}$,
it is consistent with recent experiments \cite{klapdor1,wmap}.

To make the case of an inverted spectrum of the model more transparent,
we summarize the result in this sector.
A set of the independent quantities are:
three charged fermion masses, $\theta_{13},  
\theta_{12}$ ($\simeq \theta_\nu$), $\theta_{23}$
( $\simeq \phi$  defined in (\ref{ote})), $ \Delta m_{21}^2,
\Delta  m_{23}^2$ and $\phi_\nu$,
where $\theta_q=\Delta \theta^d-\Delta \theta^u$ 
can be determined in the quark sector, and $\Delta\theta^u$
is a function of $\phi_\nu$ given in (\ref{varphi3}).
Therefore, the absolute scale for neutrino masses can be
computed from (\ref{mnu21}).  One of the absolute scales is the 
effective neutrino mass $<m_{ee}>$ in neutrinoless double $\beta$ decay,
which is shown in Fig.~4. Another prediction is  
the Dirac phase $\delta_{CP}$ defined in (\ref{dcp}),
which is plotted as a function of $\phi_\nu$ for different values of 
$\sin\theta_{13}$ in Fig.~5,
where we have assumed $\cos\theta_{23}=1/\sqrt{2}$.
As we can see from these figures,  
precise measurements of the physical parameters such as 
$\sin\theta_{13}, \delta_{CP}$ and $<m_{ee}>$  in the neutrino sector
can in principle confirm the predictions of the model.

\vspace{0.5cm}
\noindent
{\bf II Normal spectrum}

\vspace{0.5cm}
A normal hierarchal spectrum of neutrino masses with  small
$|V_{e3} |$ can be obtained if the mass parameters 
of $M_{\nu}$ (\ref{numass})
satisfy the following conditions:
\be
c_\nu^2, b^{'2}_\nu & << & b_\nu^2 , a_\nu^2~~\mbox{and}~~\varphi=
2 \Delta\theta^u\simeq 0.
\label{cond2}
\ee
So, in the lowest order approximation, we may neglect CP violating phases involved in
the unitary matrix $U_\nu$. Under the conditions (\ref{cond2}), we find that
the unitary matrix $U_\nu$ can be approximately written as
\be
U_{\nu}^{(0)}&= &\left( \begin{array}{ccc}
1  & 0 &  0
\\0  &  c_a  & s_a
\\0  & -s_a   & c_a
 \end{array}\right)
\times \left( \begin{array}{ccc}
c_{\nu}  & -s_{\nu}
&  \epsilon
\\ s_\nu &  c_\nu  & 0
\\ \epsilon c_\nu   &
\epsilon s_\nu   & 1
 \end{array}\right)
 \label{unumax4}
 \ee
where $s_\nu=\sin\theta_\nu, s_a=\sin\theta_a$ etc, and 
\be
 \epsilon &\simeq & 
\frac{R_\nu}{2}\cot\theta_a \sin 2\theta_\nu, \\
\label{epsilon}
R_\nu &=&
\frac{ ( m_{\nu_2}- m_{\nu_1})}{ m_{\nu_3}  -m_{\nu_2}\sin^2\theta_\nu
-  m_{\nu_1}\cos^2\theta_{\nu}}.
\label{rnu}
\ee
The mixing matrix $V_{\rm MNS}$ is  obtained from
$O^T_e P_e^\dag  P_{\nu} U_\nu$, where $O^T_e,  P_e$ and  $ P_\nu$
are given in (\ref{ote}), (\ref{Pe}), (\ref{Pnu}).
Note, however, that because of the condition (\ref{cond2}), 
$\arg <H_1^u> \simeq \arg <H_3^u>$ so that 
$P_{\nu} \propto {\bf 1}$. We find that 
the mixing matrix $V_{\rm MNS}$ is approximately given by
\be
V_{e2}
&\simeq & -\sin \theta_\nu,\\
V_{e3}
&\simeq  & \epsilon+
\sqrt{\frac{m_e}{m_\mu}} \left(\sqrt{\cos\phi}\sin\theta_a e^{-i2 \Delta\theta^d}-
\frac{\sin\phi}{\sqrt{\cos\phi}} \cos\theta_a e^{-i \Delta\theta^d}\right),
\label{ve32}\\
V_{\mu3} 
&\simeq & \sin\theta_a  \cos\phi e^{- 2 i \Delta \theta^d}
-\cos\theta_a\sin\phi e^{-  i \Delta \theta^d},\\
V_{\tau3}
&\simeq  &\sin\theta_a  \sin\phi e^{- 2 i \Delta \theta^d}
+\cos\theta_a\cos\phi e^{-  i \Delta \theta^d}, 
\ee
where $\Delta\theta^d$ and $\phi $ are  defined in (\ref{arg1}) and (\ref{ote}),
respectively.

The condition (\ref{cond2}) in the normal hierarchal case requires that 
$\Delta\theta^u \simeq 0$, which implies that $\theta_q \simeq \Delta\theta^d$.
We recall that  $\theta_q$ is  determined in the quark sector
 (realistic values of $\theta_q$ are between $-1.3 $ and $ -1.0$
 as it is  given in (\ref{range})). Therefore, 
 if we use the experimental information on $|V_{\mu 3}|$,
 we can relate  two parameters  $\phi$ and $\theta_a$.
 Below we assume that the mixing of the atmospheric neutrino
 is close to maximal and that $\sin\theta_a $ is positive, 
  and we regard $\sin\phi$ as independent.
  In Fig.~6 we plot $|V_{e3}|^2$ 
 against $\sin\phi$  for different values
of $R_\nu$ with $\theta_q=-1.15$.  In Fig.~7,
$\sin\delta_{CP}$ 
(defined in the standard form (\ref{dcp}) under the assumption 
$\sin\theta_{13} >0$) against $\sin\phi$
is plotted.
\begin{center}
\begin{figure}[htb]
\includegraphics*[width=0.6\textwidth]{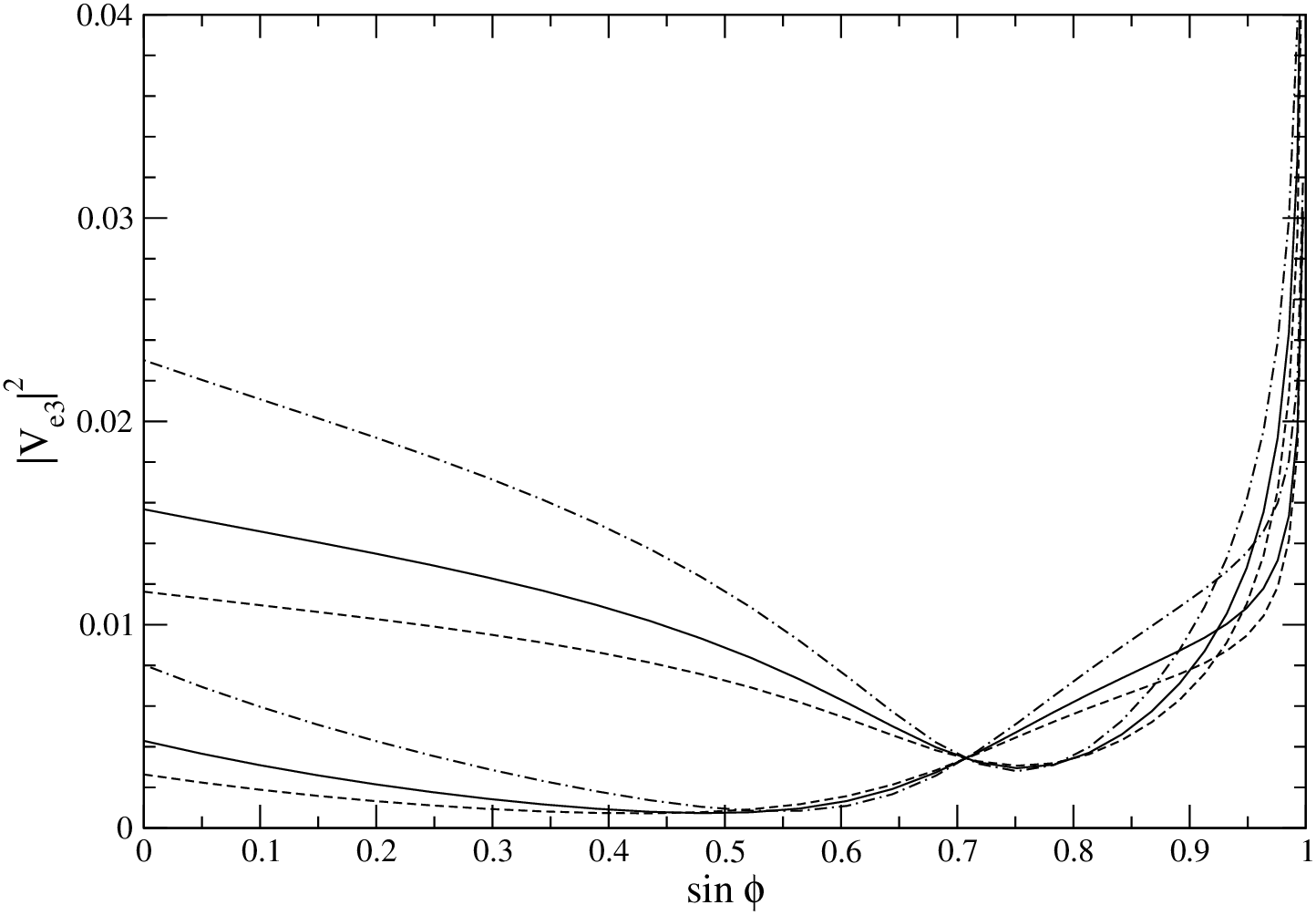}
\caption{\label{fig6}
$|V_{e3}|^2$  against $\sin\phi$ for $R_\nu= 0.15$ (doted),
$ 0.19$ (solid) and $0.25$ (dashed-doted) with $\theta_q=-1.15$
and $\sin^2\theta_{\nu}=0.3$,
where $R_\nu$ and $\theta_q$ are defined in (\ref{rnu}) and (\ref{Pq}),
respectively.
The upper (lower) lines of each pair at $\sin\phi=0$
correspond to $\sin \theta_\nu 
(\simeq \sin\theta_{12})=-(+)\sqrt{0.3}$. The neutrino
mass spectrum is normal.}
\end{figure}
\end{center}
\begin{center}
\begin{figure}[htb]
\includegraphics*[width=0.6\textwidth]{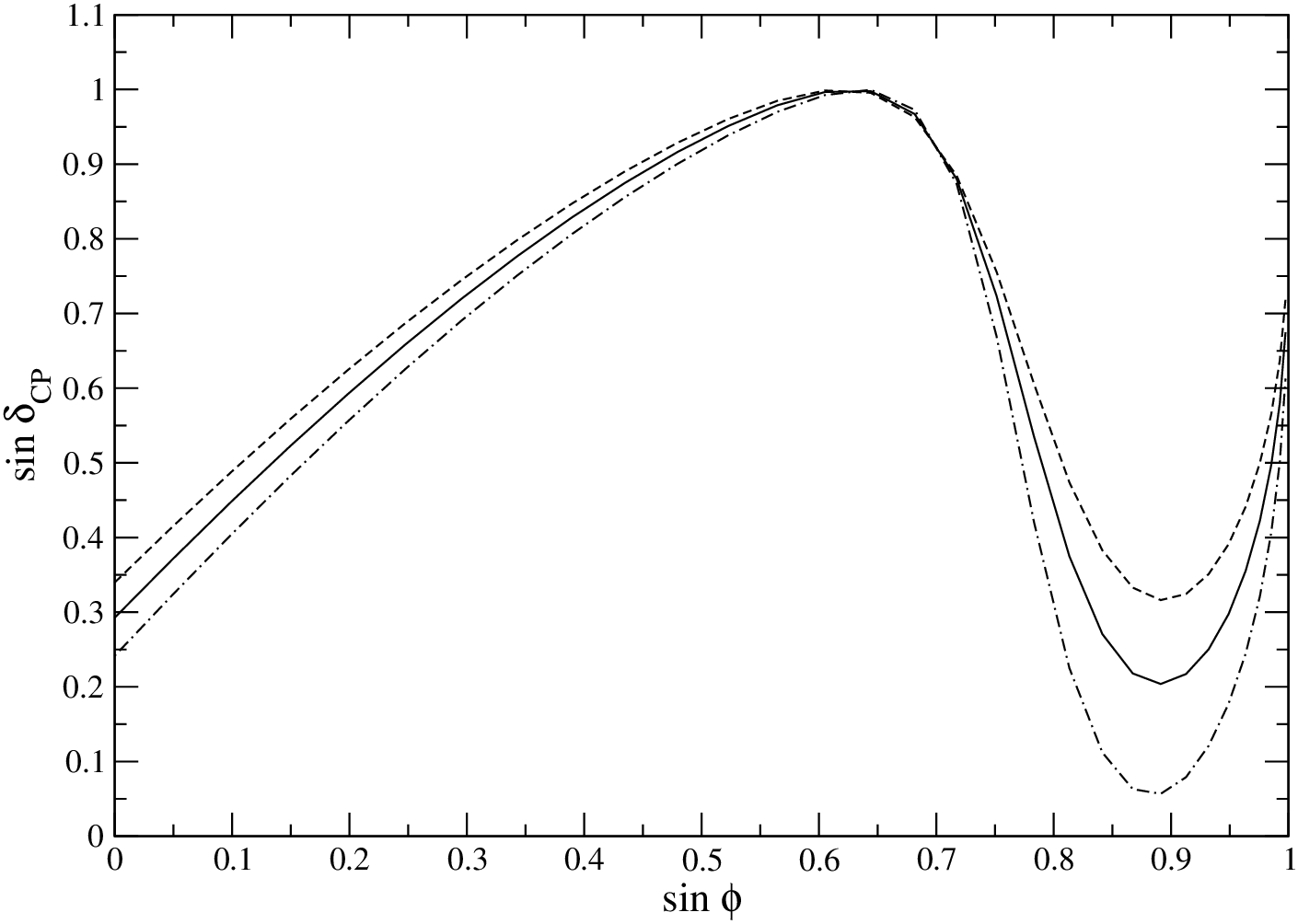}
\caption{\label{fig7}
$\sin\delta_{CP}$ against $\sin\phi$  for $R_\nu= 0.15$ (doted),
$ 0.19$ (solid) and $ 0.25$ (dashed-doted) 
with $\sin\theta_{\nu}=-\sqrt{0.3}$.
}
\end{figure}
\end{center}
\begin{center}
\begin{figure}[htb]
\includegraphics*[width=0.6\textwidth]{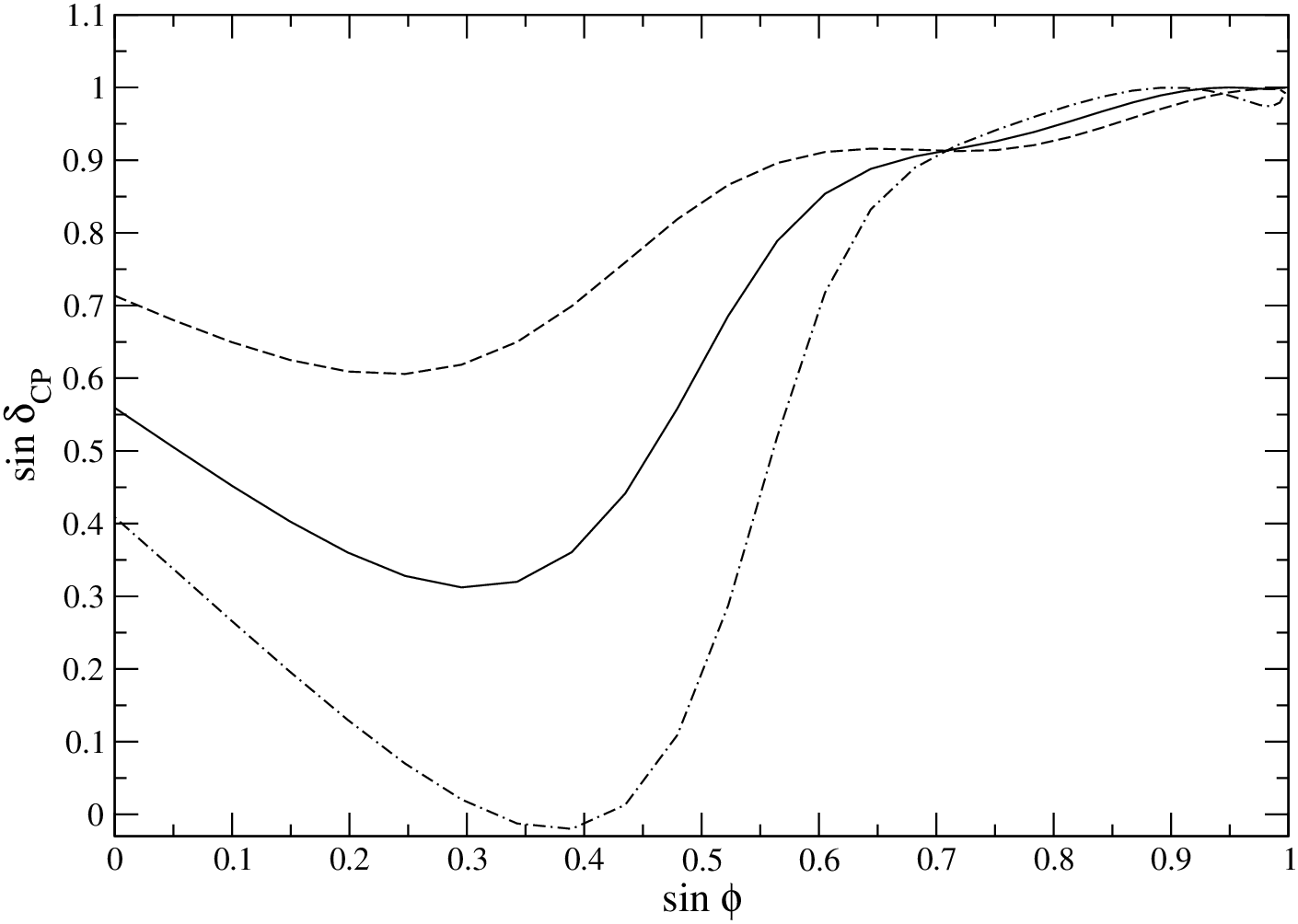}
\caption{\label{fig8}
The same as Fig.~7  for  $\sin\theta_{\nu}=+\sqrt{0.3}$.}
\end{figure}
\end{center}

As we see from Figs.~6-8, in the case of a normal 
neutrino mass spectrum , too, precise measurements of the physical parameters in the neutrino sector
can in principle confirm the predictions of the model.

\subsection{Higgs sector}
The Higgs supermutiplets
$H^{u,d}$ and $H_3^{u,d}$ belong
to ${\bf 2'}$ and ${\bf 1'''}$, respectively. 
Therefore, the singlets $H_3^{u,d}$ can not form a mass term.
So, the $Q_6$ invariant superpotential is 
\be
W_H &=&
\mu_H (H^u_1H^d_1+H^u_2 H^d_2),
\ee
which has an accidental $U(2)$ symmetry, implying
that after spontaneous symmetry breaking of the electroweak
gauge symmetry, there will appear quasi Nambu-Goldstone
supermultiplets. Moreover, the mass term for $H_3$ is absent, so that
its fermionic component will remain massless even after SUSY breaking.
So, they will conflict with experimental observations.

The origin of the so-called $\mu$ term is always a mysterious problem.
It is usually assumed that it is  related to SUSY breaking.
In the present model with $Q_6$ family symmetry,
we also worry about breaking of $Q_6$, because our assumption 
on VEVs (\ref{vev1}) means a complete breaking of $Q_6$.
It may be interesting to construct a Higgs sector, in which  
the origin of the $\mu$ term and $Q_6$ breaking
are  related to SUSY breaking.
It turns out that to construct such a sector, we need to introduce a number of
$SU(2)_L$ singlet Higgs multiplets, and also an additional
abelian discrete symmetry $Z_{12R}$, which acts as an $R$--symmetry.
$Q_6 \times Z_{12R}$ assignments are given in Table 1 and 2.

\vspace{0.5cm}
\begin{center}
\begin{tabular}{|c|c|c|c|c|c|}
\hline
 & $S$ 
& $T$ 
& $X$   & $Y$   & $Z$ 
\\ \hline
$Q_6$ &${\bf 2}$ & ${\bf 2'}$ &
${\bf 1}$ & ${\bf 1'}$ & ${\bf 1'}$
\\ \hline
$Z_{12R}$ & $2$ 
& $2$ & $-2$  
& $6$   & $0$  
\\ \hline
\end{tabular}

\footnotesize
{\bf Table 2}. $Q_6 \times Z_{12R}$ assignment 
of the $SU(3)_C \times SU(2)_L\times U(1)_Y$ singlet Higgs supermultiplets.
\end{center}

\noindent
Then we consider the following 
 most general $Q_6\times Z_{12R}$ invariant superpotential
which contains the right-handed neutrino and Higgs supermultiplets:
\be
W_H' &=& W_I+W_{II},
\label{WHp}
\ee
where the $Z_{12R}$ charge of  $W_H'$ is $-2~(\mbox{mod}~12)$, and 
\be
W_I &=& M_1(N_1^2+N_2^2) 
+\lambda_N Z  N_3^2+ \lambda_Z X(Z^2-M^2),
\label{WI}\\
W_{II}&=&
\lambda_1 X Y^2 
+\lambda_2 (S_1 S_1+S_2 S_2) Y\nn\\
& &+\kappa_1 (H_1^u S_2+H_2^u S_1) H_3^d
+\kappa_2 (H_1^d S_2+H_2^d S_1) H_3^u \nn\\
 & &+\kappa_3 \left[ -(H_1^u H_1^d-H_2^u H_2^d) T_1
+ (H_1^u H_2^d+H_2^u H_1^d) T_2 \right].
\label{WII}
\ee
We assume that $O(M_1) \sim O(M)$.
In the SUSY limit, there is a local minimum, where $\left\langle
Z \right\rangle  =M$ and all other VEVs are zero.  After SUSY
breaking, $X$  will acquire a VEV of order $M_{\rm SUSY}$,
which will induce effective $\mu$ terms for the other Higgs 
supermutiplets in (\ref{WII}).
The  charges of the fields are so chosen that the most general 
$Q_6\times Z_{12R}$ 
invariant Higgs superpotential (\ref{WHp}) 
has an accidental parity invariance:
\be
H_1^{u,d} &\leftrightarrow & H_2^{u,d}, 
S_1\leftrightarrow S_2, T_1\rightarrow -T_1,
\label{parity}
\ee
where $H_3^{u,d}, X, Y,Z$ and $T_2$ do not transform.
This enables us to choose a VEV structure given by
\be
\left\langle H_1^u \right\rangle &=& \left\langle H_2^u
\right\rangle,~ \left\langle H_1^d \right\rangle = 
\left\langle H_2^d\right\rangle,~\left\langle H_3^{u,d}\right\rangle\neq 0,\\
\left\langle S_1 \right\rangle &=&
\left\langle S_2 \right\rangle,~\left\langle T_1\right\rangle
=0,~~\left\langle T_2\right\rangle \neq 0.
\label{vev2}
\ee
Moreover, it is found that
the superpotential $W_H'$  (\ref{WHp}) induces spontaneous CP violation, when all
couplings are taken to be real.

The Higgs sector  given in this subsection 
may not be minimal. It is however clear that the Higgs sector is related to SUSY breaking,
so that there will be different modifications.
The important point is that the effective superpotential for
the $SU(2)_L$ doublet Higgs multiplets should possess
 the $H_1 \leftrightarrow H_2$ symmetry to ensure the VEV structure (\ref{vev1}).
Therefore, instead of introducing the $S,T,\dots$ singlets fields, one may
assume that
$Q_6$ is explicitly, but softly broken by dimension $3$ terms.  If we
demand that these soft terms do not break the $H_1 \leftrightarrow
H_2$ symmetry, it must be of the form
\be
W_{\rm eff} &=
&\mu_1(H_1^u H_1^d + H_2^u H_2^d)  + \mu_3H_3^u H_3^d
+ \mu_{13} (H_1^u + H_2^u) H_3^d \nn\\
& &+ \mu_{31} H_3^u (H_1^d+ H_2^d)+ \mu_{12} (H_1^u H_2^d+ H_2^uH_1^d).
\ee
Since the $\mu$ parameter results from VEVs, they may be complex.

 In models with more than one  Higgs $SU(2)_L$ doublet,
as in the case of  the present model,
tree-level FCNCs exist in the Higgs sector.
They contribute, for instance, to the mass difference 
$\Delta m_{K}$ of $K^{0}$ and $\overline{K}^{0}$,
which will give severe constraints on the model,
especially on the heavy Higgs masses.
 [See for instance  \cite{yamanaka,brown}.]
 This is one of the important questions, but 
detailed investigation of this question is
 beyond the scope of the present paper.
 So we would like to leave this problem to future work.

\section{Discussions}
Here we would like to discuss 
further features of the $Q_6$ model.

\subsection{Anomalies of discrete symmetries}
Quantum gravitational effects violate all global symmetries, while
they respect all local symmetries \cite{hawking}. 
When a local gauge group is spontaneously
broken, a certain discrete subgroup may survive \cite{krauss}, which is
respected by gravity. A necessary condition for a discrete symmetry group
to be a subgroup of a gauge group is the absence of anomalies \cite{banks,ibanez,babu4}.

We start with $Q_6$ anomalies 
\footnote{$Q_6$ anomalies are also considered in \cite{frampton1}.},
$Q_6 [SU(2)_L]^2$ and $Q_6 [SU(3)_C]^2$.
Since the anomaly in $Q_6 [U(1)_Y]^2$ depends on the normalization of $U(1)_Y$,
it does give a useful information.
Similarly, heavy fields can contribute to the $[Q_6]^3$  anomaly 
 so that it does not constrain the low-energy spectrum.
To compute anomalies, we note that 
$A_{Q_6},( A_{Q_6})^2, \dots, (A_{Q_6})^5$ correspond to  $Z_6$
rotations, and $B_{Q_6}$ to  $Z_4$ rotations.
[$A_{Q_6}$ and $B_{Q_6}$ are defined in (\ref{AD}) and (\ref{BD}).]
One can easily convince oneself that the doublets of $Q_6$ 
can not contribute to the anomalies, because
they always contain two components, one with a plus charge and the other one
with a negative charge in the same amount.
The charges of $Q_6$ singlets under ($Z_6,Z_4$) are given by
\begin{eqnarray}
 {\bf 1} &:& (0,0),~{\bf 1'} :
 (0,2),~{\bf 1''} :(3,-1), ~{\bf 1'''}: (3,1).
\end{eqnarray}
From Table 1 we can  read off  the corresponding charges
of the matter supermultiplets , and we then compute the anomaly coefficients:
\be
Z_6 [SU(2)_L]^2 &:&
2A_2=3~ (\mbox{mod}~6),~Z_6 [SU(3)_C]^2 :
2A_3=3  ~(\mbox{mod}~6),\\
Z_4 [SU(2)_L]^2 &:&
2A_2=2 ~(\mbox{mod}~ 4),~Z_4 [SU(3)_C]^2 :
2A_3=2 ~ (\mbox{mod} ~4).
\ee
$Z_{12R} [SU(2)_L]^2$ and $Z_{12R} [SU(3)_C]^2$
anomalies can also be straightforwardly computed, and we find
\be
Z_{12R} [SU(2)_L]^2 &:&
2A_2=6~ (\mbox{mod} ~12),~Z_{12R} [SU(3)_C]^2 :
2A_3=6  ~(\mbox{mod}~ 12).
\ee
Anomalies of a discrete symmetry can be cancelled
by the Green-Schwarz mechanism, if
\be
\frac{ A_2+m N/2}{k_2} &=& \frac{ A_3+m' N/2}{k_3}
\ee
is satisfied, where $k$'s ( the Kac-Moody levels ) and
$m,m'$ are integers, 
and $N$ stands for the order of $Z_N$   \cite{banks,ibanez,babu4}.
Therefore,  all the anomalies cancel
if we choose
\be
k_2 &=&  k_3,
\ee
for instance.

\subsection{$R$-parity and Baryon number violating
operators}
Just because of $Q_6$ symmetry, it
turns out that $R$-parity violating couplings are almost absent.
 Out of the 96  $R$-parity breaking cubic couplings that are allowed in the MSSM superpotential,
$Q_6$ allows only one coupling
\be
\lambda' [(L_1 Q_2 + L_2 Q_1)D_1^c + (L_1 Q_1 - L_2 Q_2) D_2^c].
\ee
Many couplings vanish because of color antisymmetry
and$ SU(2)_L$ antisymmetry.  Furthermore,  all baryon number violating 
cubic terms are
forbidden by $Q_6$ alone.
This means that there is no proton decay problem in the present 
model \footnote{Anomaly-free abelian discrete symmetries to suppress
 proton decay have been considered in \cite{babu4}.}. 
The natural value of $\lambda'$ is about
$10^{-3}$, consistent with Yukawa couplings of charged sector.
 This term will induce a neutrino mass proportional to the strange quark mass
at one loop.  The magnitude may be  of order  $10^{-2}$ eV.

\subsection{The SUSY flavor problem and phase alignment}
Since $Q_6$ is an intact symmetry at $M_{\rm SUSY}$
(except for the mass term of $N^c_3$, which
results from the VEV of $Z$ in (\ref{WI})), it is natural to assume that
the SSB sectors also respect $Q_6$ symmetry.
From the results of \cite{kobayashi1,choi1},
in which a detailed analysis on 
FCNCs and CP violations in a supersymmetric extension of
the SM with an nonabelian
discrete flavor symmetry $S_3$ has been made, we may 
expect that $Q_6$ suppresses strongly
FCNC and CP violating processes that are induced by the SSB terms.
However, the  constraints coming  from the EDM of neutron, 
electron and mercury atom \cite{gabbiani1,abel1,babu5,pdg}
are very severe.
Recent experiment on mercury atom \cite{hg} requires that
the imaginary part of the dimensionless quantity 
$(\delta^d_{11})_{LR}$ has to satisfy $|\mbox{Im}(\delta^d_{11})_{LR}|
< 6.7 \times 10^{-8} (\tilde{m}_q/100~\mbox{GeV})^2$ \cite{abel1},
where $\tilde{m}_q$ is the average of the squark masses.
Similar constraints exist for $(\delta^u_{11})_{LR}$ and $(\delta^e_{11})_{LR}$, too.
The quantity $\delta^d_{LR}$ is defined as \cite{gabbiani1}
\be
\delta^d_{LR} &=& 
\frac{U_{d}^{T}~ {\bf \tilde{m}^2}_{dLR} ~U_{d^c}}{\tilde{m}_q^2},
\ee
where ${\bf \tilde{m}^2}_{dLR}$ resulting from the A terms
is the so-called  left-right mass matrix for  the 
bosonic components 
of $D$ and $D^c$.
The unitary matrices are those that rotate the fermionic superpartners,
i.e. $U_d=P_d O_d, U_{d^c}=P_{d^c} O_{d^c}$,
where 
the phase matrices $P_d$ and $P_{d^c}$ are defined in 
 in (\ref{Pu}) and  (\ref{Puc}),  respectively.
Since we assume that  the CP phases are spontaneously induced,
and thanks to $Q_6$ symmetry 
the mass matrix ${\bf \tilde{m}^2}_{dLR}$
has  exactly the same structure as the mass matrix (\ref{nni2}), we conclude that
$(\delta^d_{11})_{LR}$ is a real number.
From the same reason, all $\delta_{LR}$ are real.
Thus, we can satisfy the most stringent constraint on 
the A terms without any fine tuning.
This is true not only at a particular energy scale, but also for the entire energy scale,
which should be compared with the case of the MSSM \cite{babu5}.

\vspace{0.5cm}
\noindent
{\large \bf Acknowledgments}\\
This work is supported by the Grants-in-Aid for Scientific Research 
from the Japan Society for the Promotion of Science
(No. 13135210).

\appendix

\section{Multiplication rules}
Here we give the multiplication rules for $D_3,D_4,D_6, D_8, Q_4$ and $ Q_8$.
The multiplication rules (without the Clebsch-Gordan coefficients) for finite groups with larger orders are
given in \cite{frampton2}.

\subsection{$D_3(S_3)$}
 The irreps of $D_3$ are ${\bf 2},
 ~{\bf 1},~{\bf 1'}$.
\be
\begin{array}{ccccccccc}
 {\bf 2}  &  \times   
&  {\bf 2}  &  =  &  {\bf 1'} 
&  +   &  {\bf 1} & + &  {\bf 2} 
\\ 
 \left(\begin{array}{c} x_1 \\ x_2  \end{array} \right)   & 
 \times    &  \left(\begin{array}{c} y_1 \\  y_2  \end{array} \right)  
&  =  &   (x_1 y_2 - x_2 y_1)   &  &
 (x_1 y_1 +x_2 y_2)   &    &
 \left(\begin{array}{c}-x_1 y_1 + x_2 y_2 \\
 x_1 y_2 +x_2 y_1  \end{array} \right) .\\ 
\end{array}\nn
  \label{multid31}
\ee

\subsection{$D_4$}
 The irreps of $D_4$ are ${\bf 2},
 ~{\bf 1},~{\bf 1'},~{\bf 1''},~{\bf 1'''}$.
 
\be
\begin{array}{ccccccc}
 {\bf 2}  &  \times   
&  {\bf 2}  &  =  &  {\bf 1} 
&  +   &  {\bf 1''} 
\\ 
 \left(\begin{array}{c}x_1 \\ x_2  \end{array} \right)   & 
 \times    &  \left(\begin{array}{c}y_1 \\  y_2   \end{array}\right)  
&  =  &   (x_1 y_1 + x_2 y_2)   &  &
 (x_1 y_1 -x_2 y_2)  \\ 
 & &  &  +  &  {\bf 1'} 
&  +   &  {\bf 1'''} 
\\ 
 &  & &  &  (x_1 y_2 - x_2 y_1)   &  &  (x_1 y_2 +x_2 y_1) . \\ 
\end{array}\nn
\label{multi41} 
\ee

\subsection{$D_6$}
 The irreps of $D_6$ are ${\bf 2}, ~ {\bf 2'},
 ~{\bf 1},~{\bf 1'},~{\bf 1''},~{\bf 1'''}$.

\be
\begin{array}{ccccccccc}
 {\bf 2}  &  \times   
&  {\bf 2}  &  =  &  {\bf 1'} 
&  +   &  {\bf 1} & + &  {\bf 2'} 
\\ 
 \left(\begin{array}{c} x_1 \\ x_2  \end{array} \right)   & 
 \times    &  \left(\begin{array}{c} y_1 \\  y_2  \end{array} \right)  
&  =  &   (x_1 y_2 - x_2 y_1)   &  &
 (x_1 y_1 +x_2 y_2)   &    &
 \left(\begin{array}{c}x_1 y_1 - x_2 y_2 \\
 x_1 y_2 +x_2 y_1  \end{array} \right) ,\\ 
\end{array}\nn
  \label{multid61}
\ee
\be
\begin{array}{ccccccccc}
 {\bf 2'}  &  \times   
&  {\bf 2'}  &  =  &  {\bf 1} 
&  +   &  {\bf 1'} & + &  {\bf 2'} 
\\ 
 \left(\begin{array}{c}a_1 \\ a_2  \end{array} \right)   & 
 \times    &  \left(\begin{array}{c}b_1 \\  b_2   \end{array}\right)  
&  =  &   (a_1 b_1 + a_2 b_2)   &  &
 (a_1 b_2 -a_2 b_1)   &    &
 \left(\begin{array}{c}-a_1 b_1 + a_2 b_2 \\ a_1 b_2 +a_2 b_1 \end{array} \right) ,\\ 
\end{array}\nn
\label{multid62} 
\ee
\be
\begin{array}{ccccccccc}
 {\bf 2}  &  \times   
&  {\bf 2'}  &  =  &  {\bf 1'''} 
&  +   &  {\bf 1''} & + &  {\bf 2} 
\\ 
 \left(\begin{array}{c}x_1 \\ x_2  \end{array} \right)   & 
 \times    &  \left(\begin{array}{c}a_1 \\  a_2 \end{array} \right)  
&  =  &   (x_1 a_2 + x_2 a_1)   &  &
 (x_1 a_1 -x_2 a_2)   &    &
 \left(\begin{array}{c}x_1 a_1 + x_2 a_2 \\ x_1 a_2 -x_2 a_1 \end{array} \right) .  \\ 
\end{array}\nn
  \label{multid63}
\ee

\subsection{$D_8$}

 The irreps of $D_8$ are ${\bf 2}, ~ {\bf 2'},~{\bf 2''},
 ~{\bf 1},~{\bf 1'},~{\bf 1''},~{\bf 1'''}$.

\be
\begin{array}{ccccccccc}
 {\bf 2}  &  \times   
&  {\bf 2}  &  =  &  {\bf 1'} 
&  +   &  {\bf 1} & + &  {\bf 2'} 
\\ 
 \left(\begin{array}{c} x_1 \\ x_2  \end{array} \right)   & 
 \times    &  \left(\begin{array}{c} y_1 \\  y_2  \end{array} \right)  
&  =  &   (x_1 y_2 - x_2 y_1)   &  &
 (x_1 y_1 +x_2 y_2)   &    &
 \left(\begin{array}{c} x_1 y_1 - x_2 y_2\\
 x_1 y_2 + x_2 y_1 \\  \end{array} \right) ,\\ 
\end{array}\nn
  \label{multi80}
\ee

\be
\begin{array}{ccccccc}
 {\bf 2'}  &  \times   
&  {\bf 2'}  &  =  &  {\bf 1} 
&  +   &  {\bf 1''} 
\\ 
 \left(\begin{array}{c}a_1 \\ a_2  \end{array} \right)   & 
 \times    &  \left(\begin{array}{c}b_1 \\  b_2   \end{array}\right)  
&  =  &   (a_1 b_1 + a_2 b_2)   &  &
 (a_1 b_1 -a_2 b_2)  \\ 
 & &  &  +  &  {\bf 1'} 
&  +   &  {\bf 1'''} 
\\ 
 &  & &  &  (a_1 b_2 - a_2 b_1)   &  &  (a_1 b_2 +a_2 b_1) , \\ 
\end{array}\nn
\label{multi81} 
\ee

\be
\begin{array}{ccccccc}
 {\bf 2}  &  \times   
&  {\bf 2'}  &  =  &  {\bf 2} 
&  +   &  {\bf 2''} 
\\ 
 \left(\begin{array}{c}x_1 \\ x_2  \end{array} \right)   & 
 \times    &  \left(\begin{array}{c}a_1 \\  a_2 \end{array} \right)  
&  =  &
 \left(\begin{array}{c}x_1 a_1 + x_2 a_2 \\ x_1 a_2 -x_2 a_1 \end{array} \right)   &    &
 \left(\begin{array}{c}x_1 a_1 - x_2 a_2 \\ x_1 a_2 +x_2 a_1 \end{array} \right) , \\
\end{array}\nn
  \label{multi83}
\ee

\be
\begin{array}{ccccccc}
 {\bf 2''}  &  \times   
&  {\bf 2'}  &  =  &  {\bf 2} 
&  +   &  {\bf 2''} 
\\ 
 \left(\begin{array}{c}t_1 \\ t_2  \end{array} \right)   & 
 \times    &  \left(\begin{array}{c}a_1 \\  a_2 \end{array} \right)  
&  =  &
 \left(\begin{array}{c}-t_1 a_1 - t_2 a_2 \\ t_1 a_2 -t_2 a_1 \end{array} \right)   &    &
 \left(\begin{array}{c}t_1 a_1 -t_2 a_2 \\ -t_1 a_2 -t_2 a_1 \end{array} \right) .  \\ 
\end{array}\nn
  \label{multi82}
\ee

\be
\begin{array}{ccccccccc}
 {\bf 2}  &  \times   
&  {\bf 2''}  &  =  &  {\bf 1''} 
&  +   &  {\bf 1'''} & + &  {\bf 2'} 
\\ 
 \left(\begin{array}{c} x_1 \\ x_2  \end{array} \right)   & 
 \times    &  \left(\begin{array}{c} t_1 \\  t_2  \end{array} \right)  
&  =  &   (x_1 t_1 - x_2 t_2)   &  &
 (x_1 t_2 +x_2 t_1)   &    &
 \left(\begin{array}{c} x_1 t_1 +x_2 t_2\\
 x_1 t_2 - x_2 t_1  \end{array} \right) ,\\ 
\end{array}\nn
  \label{multi84}
\ee

\be
\begin{array}{ccccccccc}
 {\bf 2''}  &  \times   
&  {\bf 2''}  &  =  &  {\bf 1} 
&  +   &  {\bf 1'} & + &  {\bf 2'} 
\\ 
 \left(\begin{array}{c} s_1 \\ s_2  \end{array} \right)   & 
 \times    &  \left(\begin{array}{c} t_1 \\  t_2  \end{array} \right)  
&  =  &   (s_1 t_1 + s_2 t_2)   &  &
 (s_1 t_2 -s_2 t_1)   &    &
 \left(\begin{array}{c}  s_1 t_1 - s_2 t_2\\
-s_1 t_2 - s_2 t_1  \end{array} \right) ,\\ 
\end{array}\nn
  \label{multi85}
\ee

\subsection{$Q_4$}
 The irreps of $Q_4$ are ${\bf 2},
 ~{\bf 1},~{\bf 1'},~{\bf 1''},~{\bf 1'''}$,
 where ${\bf 2}$ is pseudo-real, and
the singlets are real.

\be
\begin{array}{ccccccc}
 {\bf 2}  &  \times   
&  {\bf 2}  &  =  &  {\bf 1'} 
&  +   &  {\bf 1'''} 
\\ 
 \left(\begin{array}{c}x_1 \\ x_2  \end{array} \right)   & 
 \times    &  \left(\begin{array}{c}y_1 \\  y_2   \end{array}\right)  
&  =  &   (x_1 y_1 + x_2 y_2)   &  &
 (x_1 y_1 -x_2 y_2)  \\ 
 & &  &  +  &  {\bf 1} 
&  +   &  {\bf 1''} 
\\ 
 &  & &  &  (x_1 y_2 - x_2 y_1)   &  &  (x_1 y_2 +x_2 y_1) , \\ 
\end{array}\nn
\label{multiq41} 
\ee

\subsection{$Q_6$}
The multiplication rules for $Q_6$ are given in Sect. II.

\subsection{$Q_8$}

 The irreps of $ Q_8$  are  ${\bf 2}, ~ {\bf 2'},~{\bf 2''},
 ~{\bf 1},~{\bf 1'},~{\bf 1''},~{\bf 1'''}$ .
All  singlets and  ${\bf 2'}$  are
real representations.  
\be
\begin{array}{ccccccccc}
 {\bf 2}  &  \times   
&  {\bf 2}  &  =  &  {\bf 1} 
&  +   &  {\bf 1'} & + &  {\bf 2'} 
\\ 
 \left(\begin{array}{c} x_1 \\ x_2  \end{array} \right)   & 
 \times    &  \left(\begin{array}{c} y_1 \\  y_2  \end{array} \right)  
&  =  &   (x_1 y_2 - x_2 y_1)   &  &
 (x_1 y_1 +x_2 y_2)   &    &
 \left(\begin{array}{c} -x_1 y_2 - x_2 y_1 \\ x_1 y_1 - x_2 y_2  \end{array} \right) ,\\ 
\end{array}\nn
  \label{multiq80}
\ee

\be
\begin{array}{ccccccc}
 {\bf 2'}  &  \times   
&  {\bf 2'}  &  =  &  {\bf 1} 
&  +   &  {\bf 1''} 
\\ 
 \left(\begin{array}{c}a_1 \\ a_2  \end{array} \right)   & 
 \times    &  \left(\begin{array}{c}b_1 \\  b_2   \end{array}\right)  
&  =  &   (a_1 b_1 + a_2 b_2)   &  &
 (a_1 b_1 -a_2 b_2)  \\ 
 & &  &  +  &  {\bf 1'} 
&  +   &  {\bf 1'''} 
\\ 
 &  & &  &  (a_1 b_2 - a_2 b_1)   &  &  (a_1 b_2 +a_2 b_1) , \\ 
\end{array}\nn
\label{multiq81} 
\ee

\be
\begin{array}{ccccccc}
 {\bf 2}  &  \times   
&  {\bf 2'}  &  =  &  {\bf 2} 
&  +   &  {\bf 2''} 
\\ 
 \left(\begin{array}{c}x_1 \\ x_2  \end{array} \right)   & 
 \times    &  \left(\begin{array}{c}a_1 \\  a_2 \end{array} \right)  
&  =  &
 \left(\begin{array}{c}x_1 a_1 + x_2 a_2 \\ x_1 a_2 -x_2 a_1 \end{array} \right)   &    &
 \left(\begin{array}{c}x_1 a_1 - x_2 a_2 \\ x_1 a_2 +x_2 a_1 \end{array} \right) ,  \\ 
\end{array}\nn
  \label{multiq83}
\ee

\be
\begin{array}{ccccccc}
 {\bf 2''}  &  \times   
&  {\bf 2'}  &  =  &  {\bf 2} 
&  +   &  {\bf 2''} 
\\ 
 \left(\begin{array}{c}t_1 \\ t_2  \end{array} \right)   & 
 \times    &  \left(\begin{array}{c}a_1 \\  a_2 \end{array} \right)  
&  =  &
 \left(\begin{array}{c}-t_1 a_1 - t_2 a_2 \\ t_1 a_2 -t_2 a_1 \end{array} \right)   &    &
 \left(\begin{array}{c}t_1 a_1 -t_2 a_2 \\ -t_1 a_2 -t_2 a_1 \end{array} \right) , \\ 
\end{array}\nn
  \label{multiq82}
\ee

\be
\begin{array}{ccccccccc}
 {\bf 2}  &  \times   
&  {\bf 2''}  &  =  &  {\bf 1'''} 
&  +   &  {\bf 1''} & + &  {\bf 2'} 
\\ 
 \left(\begin{array}{c} x_1 \\ x_2  \end{array} \right)   & 
 \times    &  \left(\begin{array}{c} t_1 \\  t_2  \end{array} \right)  
&  =  &   (x_1 t_1 - x_2 t_2)   &  &
 (x_1 t_2 +x_2 t_1)   &    &
 \left(\begin{array}{c} -x_1 t_2 + x_2 t_1 \\ x_1 t_1 +x_2 t_2  \end{array} \right) ,\\ 
\end{array}\nn
  \label{multiq84}
\ee

\be
\begin{array}{ccccccccc}
 {\bf 2''}  &  \times   
&  {\bf 2''}  &  =  &  {\bf 1'} 
&  +   &  {\bf 1} & + &  {\bf 2'} 
\\ 
 \left(\begin{array}{c} s_1 \\ s_2  \end{array} \right)   & 
 \times    &  \left(\begin{array}{c} t_1 \\  t_2  \end{array} \right)  
&  =  &   (s_1 t_1 + s_2 t_2)   &  &
 (s_1 t_2 -s_2 t_1)   &    &
 \left(\begin{array}{c} s_1 t_2 + s_2 t_1 \\ s_1 t_1 - s_2 t_2  \end{array} \right) .\\ 
\end{array}\nn
  \label{multiq85}
\ee

\newcommand{\bi}{\bibitem}

\end{document}